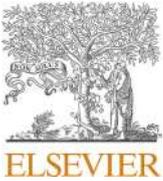

Contents lists available at ScienceDirect

# Applied Energy

journal homepage: www.elsevier.com/locate/apenergy

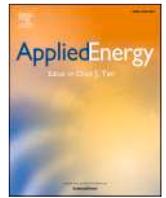

# Aperiodic two-layer energy management system for community microgrids based on blockchain strategy

Miguel Gayo-Abeleira [a,*], Carlos Santos [b], Francisco Javier Rodríguez Sánchez [a], Pedro Martín [a], José Antonio Jiménez [a], Enrique Santiso [a]

[a] Department of Electronics, University of Alcala, 28805 Alcalá de Henares, Madrid, Spain
[b] Department of Signal Theory, University of Alcala, 28805 Alcalá de Henares, Madrid, Spain

## HIGHLIGHTS

- Design of a novel two-layer aperiodic energy management strategy for community microgrids.
- Efficient penetration of distributed energy resources, maximising economic benefits.
- Integration of aperiodic local energy markets through blockchain technology.
- Execution of a distributed network topology reconfiguration through Smart Contracts.
- Implementation in low-cost devices facilitating its applicability to energy communities.

## ARTICLE INFO

## ABSTRACT

Regulatory changes in different countries regarding self-consumption and growing public concern about the environment are encouraging the establishment of community microgrids. These community microgrids integrate a large number of small-scale distributed energy resources and offers a solution to enhance power system reliability and resilience. This work proposes a geographically-based split of the community microgrids into clusters of members that tend to have similar consumption and generation profiles, mimicking the most typical layout of cities. Assuming a community microgrid divided into clusters, a two-layer architecture is developed to facilitate the greater penetration of distributed energy resources in an efficient way. The first layer, referred as the market layer, is responsible for creating local energy markets with the aim of maximising the economic benefits for community microgrid members. The second layer is responsible for the network reconfiguration, which is based on the energy balance within each cluster. This layer complies with the IEC 61850 communication standard, in order to control commercial sectionalizing and tie switches. This allows the community microgrid network to be reconfigured to minimise energy exchanges with the main grid, without requiring interaction with the distributed system operator. To implement this two-layer energy management strategy, an aperiodic market approach based on Blockchain technology, and the additional functionality offered by Smart Contracts is adopted. This embraces the concept of energy communities since it decentralizes the control and eliminates intermediaries. The use of aperiodic control techniques helps to overcome the challenges of using Blockchain technology in terms of storage, computational requirements and member privacy. The scalability and modularity of the Smart Contract-based system allow each cluster of members to be designed by tailoring the system to their specific needs. The implementation of this strategy is based on low-cost off-the-shelf devices, such as Raspberry Pi 4 Model B boards, which operate as Blockchain nodes of community microgrid members. Finally, the strategy has been validated by emulating two use cases based on the IEEE 123-node system network model highlighting the benefits of the proposal.

# 1. Introduction

There is a growing public concern over the environmental impact of the generation and distribution of the electricity. The use of conventional power plants, predominantly based on polluting fossil fuel resources, which are located hundreds of kilometres away from where the electricity is consumed, results in electricity losses and a large transmission network. This requires a considerable investment outlay. In addition, many countries, e.g., most European countries, lack of fossil fuel reserves and are therefore dependent on other countries for their energy supply [1]. In response to this public concern, during the last years, public administrations have been supporting the creation and development of Community Microgrids (CMGs) for several reasons: (i) to ensure the sustainability of the growing demand of energy; (ii) to improve the price of the energy exchanged by the members i.e., consumers, producers or prosumers; (iii) to defer or totally avoid the need to increase the transmission infrastructure; and (iv) to promote the widespread adoption of renewable energies at a domestic level as citizens' participation in renewable electricity generation is still marginal. In the European Union, the Clean Energy Package [2] provides a legislative framework to support energy communities, including those that cross property boundaries as in the case with CMGs. This framework aims to facilitate the energy transaction at the local level thereby promoting equality between the citizens and the market agents [3]. This Clean Energy Package also encourages the legislative framework transposition into national law, which will be compulsory for the Member States [4], e.g., in Spain, has been established that "power surpluses may be shared with nearby consumers located in other buildings" [5].

Compared to the traditional approach of MGs, that have been used to provide remote areas lacking electricity with a cost-effective electrical infrastructure [7], CMGs allow a group of electricity grid users to be seen as one [8]. In this new approach, the members of the CMG can exchange energy among themselves bringing clear advantages over the traditional centralized grid paradigm [9]. In this context, the main grid is used to balance the energy surplus or deficit within the CMG. Authors in [8], define a CMG as a coordinated area of the distribution grid served by one or more substations with a high rate of distributed energy resources (DERs). This coordination can be particularly complex due to the increasing inclusion of renewable energies and electric vehicles [10,11] that lead to a significant variability in the energy that users exchange with the grid.

The most common objectives [9,12] identified by Community members are: i) reducing costs and increasing benefits in their energy exchanges with the power grid; ii) achieving environmental benefits by promoting cleaner and more sustainable energy sources and reducing energy losses in transportation to the point of consumption; and iii) creating wealth and increasing grid independence, especially in the community and, as a consequence, in the country, by reducing dependence on countries with large fossil fuel reserves.

To achieve these objectives, one of the most widely researched strategies in the literature on CMGs are LEMs [13,14]. However, most of the papers present a broad perspective based on theoretical models, disregarding the practical implementation [15]. In the context of traditional MGs and pursuing the same objectives, another strategy, which is also widely researched, is based on network reconfiguration [16,17,18,19]. This strategy, to the best the author's knowledge, has never been developed in the context of CMGs, although it can be effectively implemented along with LEMs. Clearly, the combination of LEMs and network reconfiguration paves the way to achieve the goals pursued by the members comprising a CMG. To that aim, this paper proposes a CMG structure divided on geographical clusters in which members tend to have similar consumption and generation profiles (commercial cluster, industrial cluster, residential cluster, etc.) based on the most typical city layouts. Assuming this cluster-based structure, a two-layer architecture is developed. The lower layer allows LEMs to be created thereby promoting the exchange of green energy with neighbouring users. The upper layer, on the other hand, implements a network reconfiguration strategy for the entire CMG with the aim of minimising energy exchanges with the main grid.

## 1.1. Literature review

To promote energy trading and facilitate the penetration of DERs, LEMs provide a new economic and control mechanism [20,21]. This approach contrasts with the conventional market in that the prosumers exchange energy only with the distributed system operator (DSO) at a clearing price [22,23].

LEMs has been extensively reviewed in the literature. According to authors in [24], there are three different types of LEMs: (i) peer-to-peer (P2P) which allows direct transactions between users without intermediaries; (ii) transactive energy (TE) markets balance energy consumption and generation on a decentralised basis through price signals; and (iii) community self-consumption (CSC) which enables community members to trade their energy surplus. Research falling into the latter two categories can be applied to create LEM for CMGs. The full potential of LEMs could be developed by using distributed ledger technology based on blockchain (BC). Several works in the literature have employed BC to create energy markets [25]. In [26] it has been demonstrated that BC ensures that false data insertion is avoided in both energy exchanges and energy management systems. BC has a positive impact on obtaining a fair price for the exchanged energy among users [27], reducing dependence on the main grid [27] and helping to avoid peaks in the energy generation and demand curves [28].

In [26] and [29] BC implementations are carried out, but without taking into account that this implementation must be based on low-cost equipment and without analysing the resources that are consumed and that will be consumed in the long term. According to [15], most research in the field fails to tackle the practical challenges LEM poses. For instance, the ever-growing database due to the rapidly increasing requirements for storage capacity for the practical implementation of BC technology on a large scale, has been largely neglected in the literature [30].

Although in most LEM papers the energy to be exchanged by each member is known in advance, none of them take advantage of this information to adapt the network to different situations. This would be possible in future power distribution networks because sectionalizing and tie switches will have bi-directional communication and control [31]. The use of sectionalizing and tie switches already present in CMGs make network reconfiguration one of the enabling technologies to operate them in an efficient and optimal way [16]. Reconfiguration is a challenging task due to the uncertainty and intermittency of energy generation from renewable DERs [32].

While there is extensive literature on network reconfiguration applied to MGs, not much attention has been paid to network reconfiguration in CMGs. Since MGs and CMGs share common goals and features, those approaches developed for MGs can be applied to CMGs. In [33] and [34] particle swarm optimisation algorithms that attempt to minimise energy losses are presented. In [35] a reconfiguration algorithm aiming at minimising the cost of operation by considering the available DERs is described. Other works aim to maintain the frequency [36] and the voltage within the admissible ranges [37]. In [38],[39] and [40] the goal is to guarantee and restore the service as quickly as possible in the event of a failure.

From the reviewed literature on network reconfiguration, three research gaps have been found. Firstly, although there is discussion as to the future distribution network will have bi-directional communication with the physical elements [31], as far as the authors are aware, no papers have been published that use communication standards to interact with the physical elements for reconfiguration purposes. Secondly, the fact that reconfiguration can be carried out in a decentralised manner is not mentioned. Most research works only discuss the reconfiguration algorithm without providing implementation details, whereas those that present the practical implementation, focus on the system which implements it, in terms of either a microgrid's central controller [37] or the DSO [23]. As far as the DSO is concerned, the reconfiguration algorithm does not allow CMGs to reach their full potential, as they are unable to control energy flows for their own benefit. As for the microgrid's central controller, there is a single point of failure. Thirdly, all the analysed works dealing with network reconfiguration execute their algorithms periodically, such as those in [23 17]. However, the variability in the generation and demand due to stochastic nature of the renewable resources and the consumption profile, results in either sub-optimal system performance or higher than necessary resource consumption in terms of computation, data storage and message transmission.

### 1.2. Main contributions

This work proposes a geographically based split of the CMG into clusters where members tend to have similar consumption and generation profiles, reflecting the most typical layout of cities. Following a decentralised approach and using BC-based Smart Contracts (SCs), the system runs on the devices of all the community members. The system is designed with the characteristic of modularity in mind, i.e., algorithms of LEM and network reconfiguration existing in the literature can be implemented according to the characteristics of each CMG.

The purpose of dividing the CMG into different geographical clusters is threefold: (i) to deal with the ever-growing database (DB) challenge; (ii) to provide additional security to the personal information of the CMG members; and (iii) to minimise the computational requirements of the devices. Based on the division into clusters, a two-level-based BC strategy is proposed which allows that only the members of each cluster can store the data of those members in the same cluster, disregarding the data from other clusters. This strategy responds to the ever-growing DB challenge and provides additional security by reducing the number of users storing sensitive data.

Furthermore, in this work, an aperiodic strategy [30] is proposed to significantly reduce the amount of data to be stored for an effective operation. This strategy contrasts with the traditional LEMs and reconfiguration algorithms, which are executed when a certain fixed time elapses (daily, hourly, every fifteen minutes, etc…). The aperiodic approach gives very similar performance to that executed with a small period, and considerably reduces the data storage requirements and lessens the stress to the switching elements the reconfiguration layer interacts with.

The practical implementation of the energy management strategy is supported by the inclusion of the IEC 61850 [41] standard, which is considered the most promising communication standard in MGs [42]. This standard has been widely adopted in the MG environment [43]. Therefore, sectionalising and tie switches can be controlled by IEC 61850-based messages. Furthermore, a common configuration file offers plug and play features [44]. Finally, a test bench based on the Raspberry Pi 4 Model B [45] has been used to demonstrate the feasibility of implementing the strategy in off-the-shelf devices with not very advanced features.

In summary, the main contributions of this paper are as follows:

1) To the best of the authors' knowledge, is the first time that SCs, modelled on the IEC 61850 communications standard, have been used to implement a network reconfiguration algorithm, allowing CMGs to configure the network for their own benefit.
2) In the context of LEMs, a two-layer strategy is proposed and developed to overcome the challenges of utilizing BC technology in terms of storage and computational demand while preserving the privacy of the user data.
3) The reduction of the storage and computational demand is further achieved by using aperiodic control techniques. As a result, low-cost off-the-shelf devices can be used to implement the system, thereby increasing the return of investment.

4) The scalability and modularity of the SC-based system allows each cluster to be designed by adapting the system to their specific needs.

The remainder of this paper is organized as follows. Section 2 deals with the developed SC-based strategy for managing the CMG and describes the underlying technologies which enable the CMG to effectively operate. Section 3 describes the circuit on which the tests are carried out, the LEMs and the network reconfiguration algorithm included in the proposed experimental framework. Specific energy market and reconfiguration algorithms are presented in section 4. Section 5 presents the results of using the LEMs and reconfiguration algorithm designed in this work for two use cases based on the day of the week: weekday and weekend day. This section also tests the strengthens of using aperiodic techniques. Some conclusions are drawn, and future works are discussed in section 6.

## 2. Energy management approach

This section introduces the proposed energy management strategy for CMGs. The motivation behind this approach is based on the new paradigm of energy grids, created by the introduction of distributed generation and the grid digitalization. These changes are promoting worldwide a growing interest of public administrations and citizens in the creation of CMGs with the aim of reducing costs, encouraging the integration of renewable resources into the electricity grid and generating wealth [9,12]. Fig. 1 depicts the evolution of the conventional distribution network towards a CMG-based deployment. Traditionally the grid has followed a centralized approach (Fig. 1 a) that has demonstrated some well-known vulnerabilities to failures caused by natural disasters or unusual phenomena [46]. A plausible scenario consists in transforming the conventional distribution network into MGs [47] (Fig. 1 b), considering that the resulting MGs comprise users with similar profiles. This can be used to build CMGs, which try to maximise the profit for all users. The architecture devised and presented in this paper embraces this principle. Those CMGs in geographic proximity, usually are networked with the aim of reducing the energy costs and enhancing reliability and resilience [48,49,50], resulting in a mesh grid (Fig. 1 c). The merging of different CMGs results in a larger CMG, in which the original CMGs are called clusters. This facilitates the network reconfiguration to address the problems caused by faults and promotes independency from the main grid, minimizing energy exchanges.

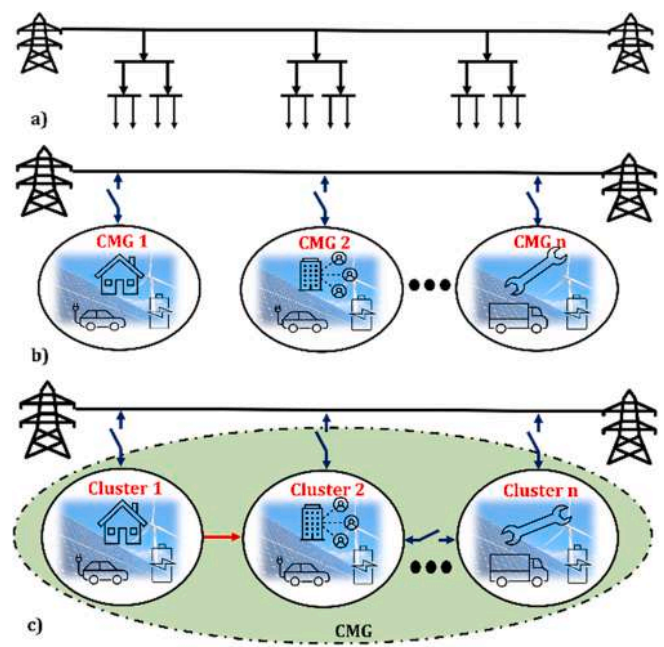

**Fig. 1.** a) Conventional distribution network. b) Distribution network split into n CMGs. c) CMG divided into n clusters with the possibility to be linked to those in geographical proximity.

Each cluster operates as an independent entity. Clusters based on residential-scale microgeneration (mainly PV installations) tend to have an energy surplus during the day and peak demand in the evening, whereas clusters composed of members from commercial or office areas have their peak consumption during the daytime. Taking advantage of these differences in profiles, energy sharing would yield considerable economic benefit to CMG [51]. It is worth mentioning that those clusters which share energy, also enhance the stability of the main grid on account of reducing the solar energy injected to the grid, which could cause problems such as overvoltages [52].

Each cluster usually has a point of common coupling (PCC) to connect to the main grid, apart from sectionalizing and tie switches used to interconnect different clusters. The approach proposed in this paper goes one step further by controlling these connection points with an Energy Management System (EMS) implemented as a SC to aggregate several clusters to create groups of clusters which can be connected to the main grid by none, one, or several PCCs. The primary aim of this proposed reconfiguration strategy is to minimise the energy exchange between the CMG and the main grid.

The proposed EMS is based on a two-layer control strategy (market layer and reconfiguration layer) as can be seen in Fig. 2. The market layer in turn consists of two sublayers: the lower market layer and the upper market layer. The lower market layer, also called cluster energy market, is used to create a LEM for each cluster and enables the first energy exchange between members within the cluster, whereas the upper market layer, so-called CMG energy market, addresses the energy exchanges between clusters once the preliminary stage of the cluster energy market has taken place. The reconfiguration layer, on the other hand, interacts with the physical elements (sectionalizing and tie switches) to establish, at any time, the topology that minimizes the amount of energy exchanged between the CMG and the main grid.

The EMS implementation relies on a business-based permissioned BC network [53]. Permissioned BC networks feature higher speed and lower power consumption compared to permissionless networks, which make them more suitable for MG applications [54]. It is also important to underline that only registered users can access the network, which significantly reduces the number of possible malicious attacks. As all users have to be registered to be members of the CMG, less secure consensus algorithms than those used in permissionless networks can be used. In this permissioned BC network, each cluster plays the role of a company, which only shares global cluster information with the CMG, whereas the details of each member are not revealed. The clusters are composed of several members. To provide SC-based energy management, each member participating in the CMG has a device that creates a BC node. This node communicates with the BC nodes of the other members.

To create the two-layer structure, a two-level blockchain has been employed. The lower BC level is associated with the lower market layer i.e., each cluster has its own BC responsible for establishing the cluster energy markets. The upper BC level, on the other hand, corresponds to the upper market layer and the reconfiguration layer i.e., creates a BC for the entire CMG. As a result, the DB is organized in two levels: the DB of each cluster and the CMG DB. Regarding the latter, confidentiality is not an issue since the CMG DB does not contain any personal information on identities, energy consumption or generation profiles. The members submit energy offers which are stored in their cluster DB. An SC extracts their cluster energy requirement data from the cluster DB and adds it to the CMG DB. Each member node holds a replication not only of the DB pertaining to the cluster they belong to, but also of the CMG DB. Again, confidentiality is not compromised since the member can only access their own data and the rest of the information is encrypted. To sum up, the two levels within the market layer provide extra protection for the members' personal data, as not only is the data encrypted but also the members who do not belong to the cluster, have no access to the data.

The purpose of this two-level-based approach is threefold. Firstly, to strengthen the security of the member data. Secondly, the data storage requirements are reduced. Finally, other clusters can be joined to form larger CMGs without negatively affecting the needs of other already installed equipment for the rest of member nodes, thereby increasing the architecture scalability.

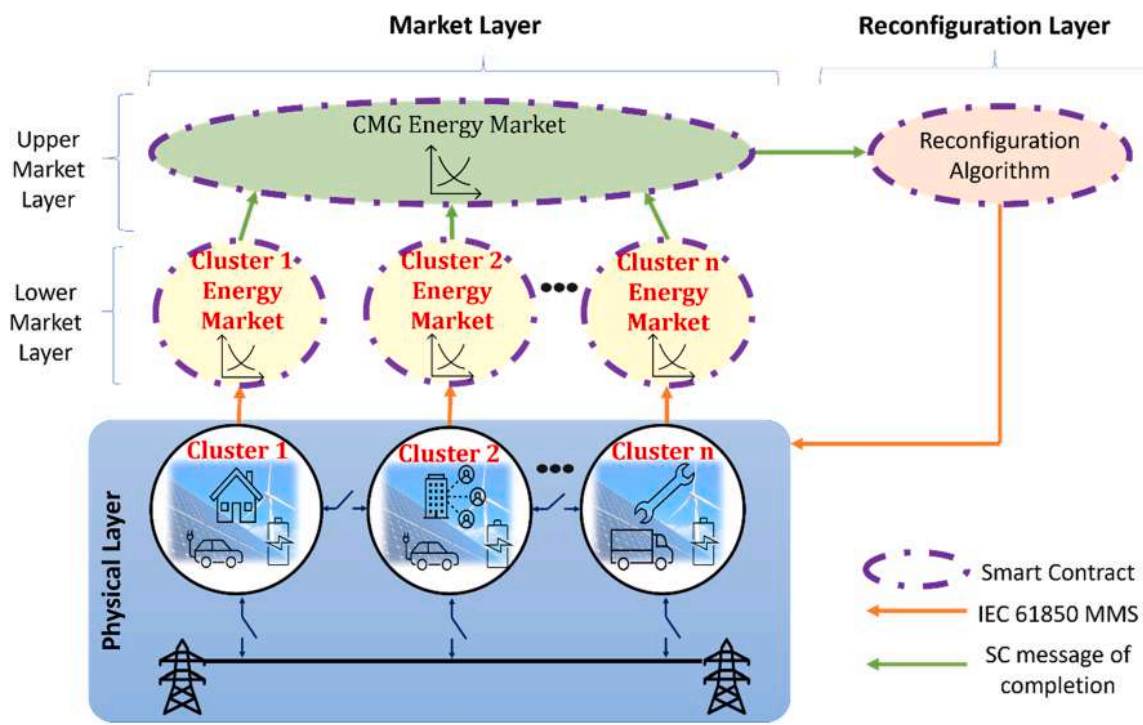

**Fig. 2.** Proposed two-layer EMS for CMGs based on SCs.

In addition to BC-based smart contracts, IEC 61850 standard is the underlying technology that allows this multilayer architecture to operate properly and securely. IEC 61850 is required to read, in a decentralized way, the smart meters located at the PCC of each cluster, those smart meters between clusters and the smart meters of each member. It is also used to send the control commands to the sectionalizing and tie switches. By ensuring that the elements that make up the CMG (smart meters, switches, …), comply with the standard, the distributed control and the reading process of their electrical parameters are guaranteed from the blockchain as proved in [30].

Likewise, the reconfiguration layer interacts with the physical elements of the CMG through IEC 61850-based Manufacturing Message Specification (MMS) protocol. The MMS messages are executed via an SC, which can be used to minimize the energy losses, to reduce the power injection to the main grid and to encourage the energy exchange among clusters to name but a few purposes.

Both the reconfiguration and the market layers do not operate independently. To establish the optimal topology, the reconfiguration layer utilizes the results obtained by the market layer when the CMG energy market is executed. In this way the future scenario for the CMG is outlined and the outcome can be anticipated.

### 2.1. Smart contracts

The proposal presented in this paper utilises two types of SCs. The first type of SC runs on each cluster comprising the CMG whereas the second one is deployed for the CMG as a whole. This provides modularity to the system, which promotes flexibility and scalability since the functions in each cluster are implemented to satisfy the cluster requirements. The SCs and the corresponding functions are:

- Cluster SC: Each cluster has its own Cluster SC. This SC is executed on the devices of the members of the cluster. This SC includes three functions:
  - Insert User Bid (IUB): This function is used for each CMG member to place its bid within the blockchain. The function is executed by the member's device.
  - Cluster Energy Market (CEM): The CEM can be specified according to the preferences or objectives of the corresponding cluster, that can differ from those of other clusters. Regardless of the energy market mechanism, the output of this function is always an energy profile in terms of the amount of energy the clusters are willing to sell or purchase in the CMG market described below. Once the cluster members have placed their bids, the execution of the CEM function is automatically triggered.
  - Read user meters (RUM): this function reads, from the member meters, the amount of energy that has been exchanged with the grid since the last execution of the function and stores that information in an immutable way in the cluster DB to be able to bill for the actual energy exchanged. Its execution is automatically triggered every time the execution of the CEM function is finished. In addition, when the function remains idle for a long period of time, it is executed autonomously.
- Community SC: all members of the CMG have this SC, which is executed on the devices of the members of the CMG. This SC includes four functions:
  - Insert Cluster Bid (ICB): This function is used by each cluster to enter the result of the execution of its CEM function on the blockchain. The execution of the ICB function is automatically triggered every time the CEM function has finished.
  - CMG Energy Market (CMGEM): The energy profiles generated by the CEM function of the cluster SCs, are the input of the CMGEM function of the community SC. This function determines the energy transactions among clusters and those with the main grid. Its execution is automatically triggered when it detects that all clusters have placed their bids.

- Network Reconfiguration (NR): This function determines the optimal network configuration and, in a distributed way, sends the commands to the switches through the IEC 61850 communication protocol to reconfigure the network [30]. As an example, an algorithm for the network reconfiguration is explained in section 3.3. This function is automatically triggered when the execution of the CMGEM function has finished.
- Read Community Meters (RCM): this function reads the meters located at the PCCs and the meters between clusters, so that the information pertaining to the actual energy exchanged and the clusters involved, is reflected in the CMG DB, in an immutable way. This function is automatically triggered when the execution of the NR function has finished.

In contrast to other implementations found in other works in the literature that send the CMGEM results to the DSO to reconfigure the network [23], in this work the results are only sent to the DSO for informational purposes, so that the DSO is aware of the amount of energy that the CMG will exchange with the main grid thereby preventing possible outages. The network reconfiguration algorithm is implemented by using the NR (Network Reconfiguration) function of the community SC which is executed in response to a signal indicating that the CMGEM results are ready. The Community SC could involve the DSO. In this way, the results would be interpreted to seek optimal performance not only for the CMG members, but also for the main grid.

The SC-based implementation of the EMS has a major advantage in terms of its improvement capacity, that is to say, if a cluster considers that a LEM mechanism better adapts to their characteristics with respect to the one already implemented, the latter is modified. To do so, each cluster member must adhere to an agreement, otherwise the old SC will still be in force. This behaviour is also displayed in the context of the CMG energy market and the network reconfiguration algorithm, although in this case there must be a unanimous consensus on the SC among all members of the CMG.

### 2.2. Aperiodic market triggering techniques

As previously stated, BC technology relies on an ever-growing distributed DB, which is updated by appending blocks containing new information. As a result, the amount of information to be stored and transmitted becomes one of the most critical parameters when it comes to designing the hardware system to implement the BC. With this in mind, this work proposes an implementation of aperiodic LEMs to reduce the amount of information to be processed and stored by the DB.

To meet this challenge and reduce the communication within the CMG, two different aperiodic LEMs are designed, which are implemented by the two best-known aperiodic techniques: the Send on Delta (SoD) and Send on Area (SoA) techniques [55].

The simplest technique is SoD. This method focuses on the difference between the member's last power measurement sent to the CMG and their current measurement. Thus, it is possible to locally check for each member when it is necessary to update the cluster energy market due to a significant power deviation in one of the members within the CMG. The power deviation which is considered as a significant change is set by the designer through the threshold $\Delta_{SoD}$. The expression to be evaluated for each member is as follows:

$$t_{i,k} = \min\{t > t_{i,k-1} \big| \big| Power_i(t) - Power_i(t_{i,k-1}) \big| \geq \Delta_{SoD}\} (1).$$

with:

$$Power_i(t) = PV_i(t) - Demand_i(t), (2).$$

where $t_{i,k}$, represents the instant when a member $i$ has a sufficiently significant change, $Demand_i$ is the load demanded by member $i$, $PV_i$ is photovoltaic power generated by member $i$, and $Power_i$ indicates the power balance of member $i$.

As for SoA, the error is integrated over time to ensure that it does not exceed a certain energy amount. This threshold is set by the designer and represented by the parameter $\Delta_{SoA}$. The member will trigger the aperiodic market whenever the following expression is true:

$$t_{i,k} = \min\left\{ t > t_{i,k-1} \,\middle|\, \int_{t_{i,k-1}}^{t} Energy_i(t) - Energy_i(t_{i,k-1}) dt \geq \Delta_{SoA} \right\} (3).$$

with:

$$Energy_i(t) = \int_{t}^{t+\Delta_t} Power_i(t) \, dt,$$

$$\Delta_t = 60s, (4).$$

where $Energy_i(t)$ is the last-minute energy of the member $i$ with respect to instant $t$ in $kWh$ and $\Delta_t$, is the integration time.

The role these aperiodic techniques play in the cluster and CMG energy markets is essential. In all cases, a trigger event at the local level of any member entails a new update of the cluster energy market. Thus, the market is always distributed and locally executed to comply with the decentralized nature of the BC technology.

In the case of a simultaneous update, a local event triggers the update process whereby all the members send their new power measurement to the market. This slightly decreases the total number of CMG updates and slightly minimizes the estimation error at the expense of a higher overhead on the communication channel.

### 2.3. Aperiodic execution flow

Fig. 3 depicts the execution flow of the different SCs. In the figure, the processes individually executed by each member and those carried out by each of the clusters are represented in yellow and green, respectively, whereas the processes of the CMG are depicted in red. Each cluster works independently. However, if one of the clusters has exceed the pre-established threshold, the CMGEM (CMG Energy Market) begins.

The aperiodic execution flow starts when a member exchanges an amount of energy which differs from that agreed to be exchanged, as can be seen in Fig. 3. This triggers the execution of the CEM function of the cluster SC to which the member belongs. If the result exceeds the aperiodic activation threshold (see section 2.2), the execution of the CMGEM function of the Community SC is triggered. For this purpose, a signal is sent to each cluster to start the execution of their corresponding

CEM. Once the CMGEM function has received the outcome of all the CEM functions, the market is cleared and the quantity of energy that each cluster has to exchange is determined. The network is reconfigured based on the results of the CMGEM function by executing the NR function.

The RUM function of each cluster and the RCM function are executed whenever the NR function finishes its execution. The aim of these two functions is to record the data generated by the actual energy exchanges in an immutable BC DB so that the members can be billed by the actual energy exchanged instead of by the market results. This allows a fair allocation of costs resulting from losses to be made. When a new relevant change is detected, the whole process starts again. However, if no no significant change has occurred over a long period of time, the RUM and RCM functions are executed to record the status of all CMG meters.

## 3. Experimental framework

To assess the viability of the strategy introduced above, a slightly modified IEEE 123-node system has been used [56]. In this circuit a CMG with 114 members is simulated and, for each member, different energy generation and consumption profiles based on consumption areas are created. Several simple LEM models and a simple reconfiguration algorithm are presented. The blockchain framework and the test bench on which it is implemented, for validation purposes, are also described. Although specific market models and a specific reconfiguration algorithm are presented, one of the advantages of the strategy developed in this paper is that it is easily adaptable to the different requirements of each CMG and its comprising clusters which require more challenging strategies.

### 3.1. IEEE 123 node test

To make the system behaviour under consideration as realistic as possible and comparable with other proposals, a slightly modified IEEE 123-node system has been used. This circuit, which represents a distribution network, has several controllable circuit breakers that allow the different zones to be isolated. In this work, each of these zones has been considered as a cluster of users. The proposed modification of the standard IEEE 123-node system only involves introducing a PCC in each of the clusters in order to be isolated from the rest. As an exceptional case, cluster 2 is left without connection to the main grid to demonstrate that the system continues to comply with the quality requirements of the grid, also improving the efficiency even though one of the PCCs is not available. Therefore, this clusters have the capability for connecting or disconnecting to the main grid and to the rest of the clusters that comprise the CMG as shown in Fig. 4.

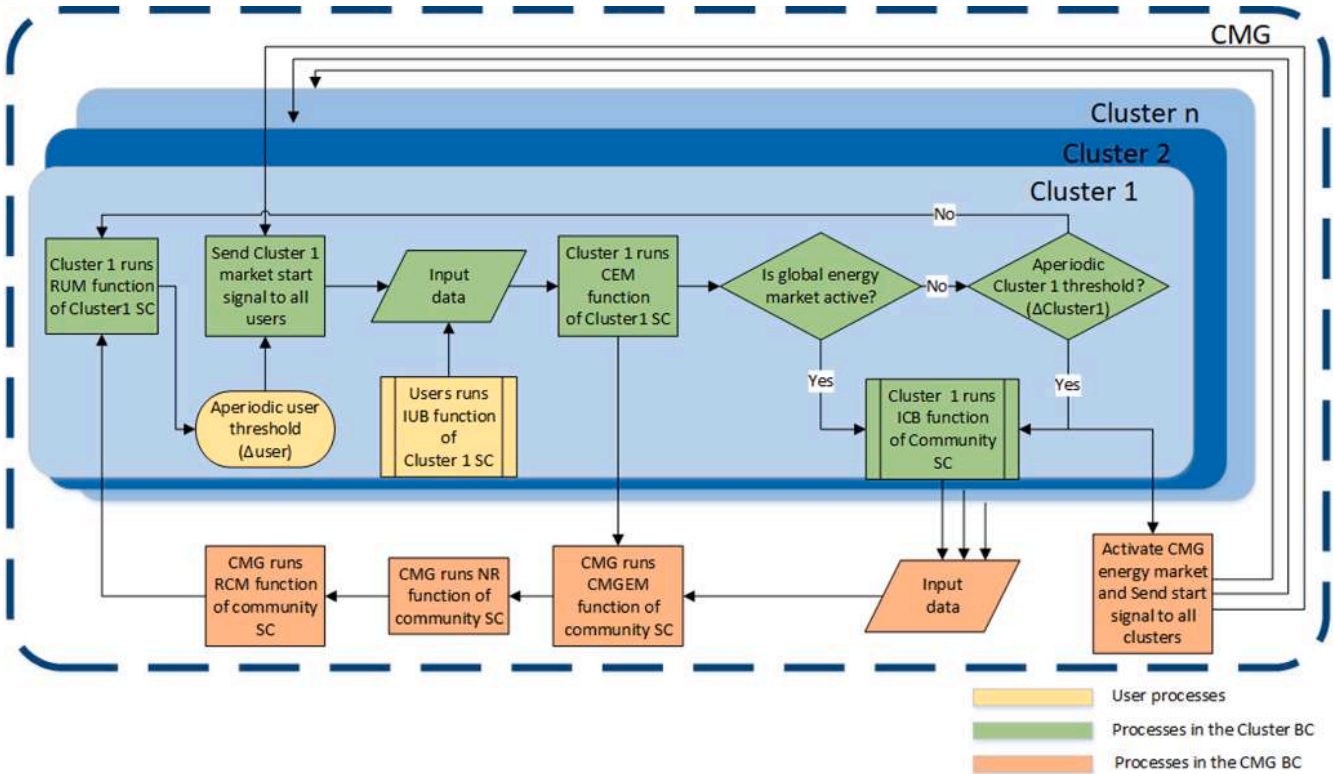

**Fig. 3.** Flowchart showing the function of the SCs that control the operation of the energy exchanges of the CMG.

The lower level of this simulation model corresponds to a current generator representing the behaviour of each node. The reference of each generator is provided by an external file which contains the energy exchange profiles with the CMG of each node as a function of time. This file is generated from the consumption and generation profiles for each node comprising the CMG. The generation and consumption data as a function of time have been built up from the database in [57]. Different roles have been established for each of the 5 clusters based on the typical layout of a city, resulting in the profiles depicted in Fig. 5: (1) Industrial cluster with 40 % of solar penetration and a load increase on working days, in red colour; (2) Commercial cluster without generation and a load increase during business hours, in blue colour; (3) University campus with 30 % of solar penetration and a load increase during school hours, in black colour; (4) single-unit residential cluster with 60 % of solar penetration and the original DB consumption profiles have been used, in cyan colour; and (5) multiple-unit residential cluster with two nodes with PV power generation and several original DB consumption profiles aggregated as the consumption profile at each node, in yellow colour. To establish different roles for the nodes, in accordance with the cluster to which they belong, generation and consumption profiles have been modified to adapt them to the orders of magnitude and schedules corresponding to their cluster. For those nodes with a generation role, the same profile based on photovoltaic generation on a sunny day, has been used. However, the magnitude of the generation profile can vary depending on the cluster to which the node belongs to.

## 3.2. Energy markets

As stated above, the energy markets are divided into two distinct stages. During the first stage the cluster energy market takes place in each cluster followed by a second stage based on the CMG energy market between clusters. The latter can only be executed when the former has finished in every cluster. The cluster energy market is implemented following the strategy presented in [30], in which the members send offers to the market. The offers essentially consist of predictions of the energy the members are expected to exchange with the main grid, from the moment the offer is delivered, these offers also include the energy price.

Once the cluster energy market receives the offers from all members, it matches some of the offers obtaining, as a result, the amount of energy the members will exchange among them and the energy price. Price is limited to the price of energy exchanges with the main grid as long as the main grid is available. The matching of offers is done by ordering all offers by price. Once ordered, the smaller quantity between the total energy offered to buy and to sell is chosen, which is automatically approved for exchange. From the larger quantity, offers with a lower price are approved to be exchanged up to the quantity approved in the previous step. The unmatched offers within each cluster energy market can be considered as the surplus or deficit of energy that can be injected to or drawn from the main grid, respectively.

To participate in the CMG energy market, each cluster generates energy purchase and sale curves from the unmatched bids in its cluster energy market. Each cluster sends these curves to the energy market of the CMG, which matches them with each other. The unmatched offers in the CMG energy market are considered as energy to be exchanged with the main grid.

## 3.3. Reconfiguration algorithm

The network reconfiguration algorithm has as input the surplus or deficit amount of energy that each cluster has after the CMG energy market has been realised, for this reason, it is executed every time the aperiodic CMG energy market is conducted because it indicates that the electricity profiles have changed as shown in Fig. 3. Its objective is to minimize the energy exchanges with the main grid while ensuring that the voltage and frequency ranges are within allowable limits by imposing technical constraints. To that end, the proposed recursive algorithm (shown in Fig. 6 as a flowchart) is executed according to the following steps:

a) **Loop to join clusters:** clusters with energy surplus, which must be injected to the grid, are joined with those with energy deficit, which must be drawn from the grid. This unification process is only feasible when there are energy paths among these clusters. If the capacity of the energy link between the two combined clusters is enough, these clusters represent a new entity for the next algorithm iteration. However, when the capacity of the link is lower than the power the clusters would individually deliver to or draw from the main grid, the allowable power limit is exchanged. In the next algorithm iteration, the involved clusters are considered as independent entities with the updated energy balance (Fig. 6a).

b) **Loop to join clusters without PCC:** If any of the resulting clusters of clusters do not have any link to the main grid available, due to either there is no direct path or a failure, the corresponding cluster is joined to the group of clusters with the highest exchange capacity available with the main grid to that of the considered cluster as can be seen in Fig. 6b.

c) **Loop to link clusters to the main grid:** the third and final stage of the reconfiguration algorithm determine what switches located at the PCCs are set to ON state so that the energy exchanges with the main grid can take place. For each group of clusters, the original cluster which should draw the most amount of energy is connected to the main grid, minimizing the energy injected to the main grid in an undesirable way. When the capacity of the energy link cannot exchange the corresponding energy of the group of clusters, the original cluster which should draw the second largest amount of energy is connected to the main grid, and so on (Fig. 6c).

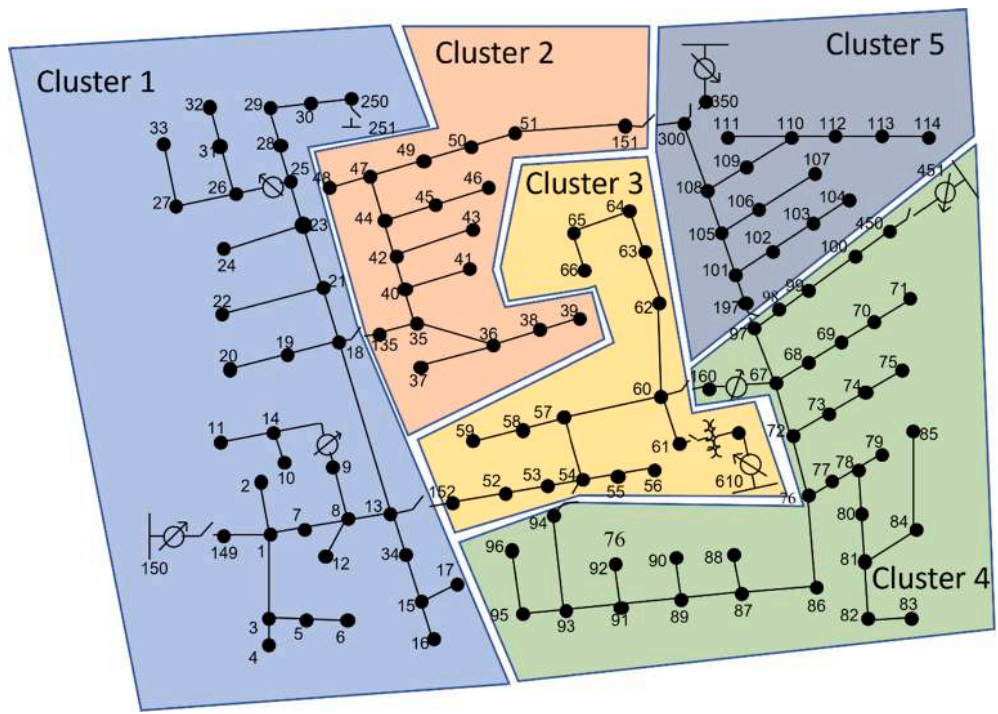

**Fig. 4.** IEEE 123 node test feeder slightly modified to create five independent clusters with possibility of connection to adjacent cluster or to the main grid.

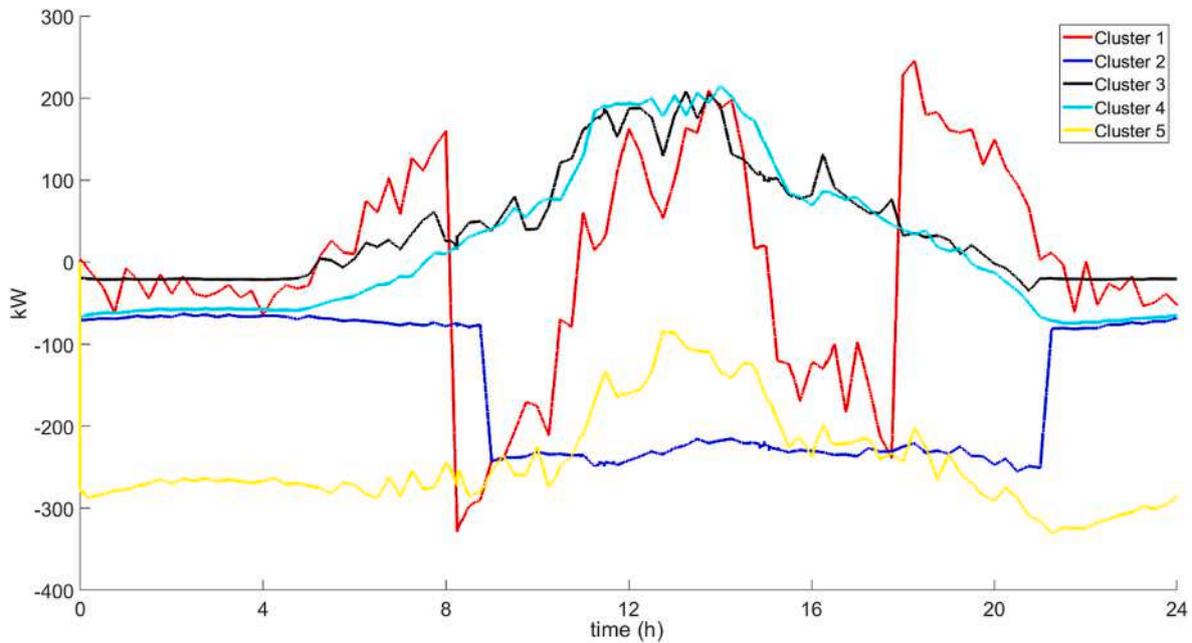

**Fig. 5.** Load profiles of each cluster for the day under study. Industrial cluster in red colour, commercial cluster in blue, university campus in black colour, single-unit residential cluster in cyan and multiple-unit residential cluster in yellow. (For interpretation of the references to colour in this figure legend, the reader is referred to the web version of this article.)

Fig. 7 shows an example of how the network reconfiguration algorithm works. Fig. 7 a) depicts a cluster with a positive power balance of 10 kW whereas the other cluster has a negative power balance of 50 kW, which means that the first cluster can deliver 10 kW and the second one could draw 50 KW. The algorithm combines both clusters and the resulting cluster has a negative power balance of 40 kW, which must be drawn from the main grid through the power link of that component cluster, i.e., one of the clusters forming the cluster, with the most negative balance. Fig. 7 b) shows the algorithm outcome when there is a current constraint which restricts the energy exchange on account of its magnitude. A cluster has a positive power balance, and the other cluster has a negative energy balance. In both cases the energy balance is greater than the link capacity and therefore both clusters must be connected to the main grid to keep their energy balance to zero. The link between the two clusters is first considered to be saturated and then the connections to the main network of each of the clusters that still has a non-zero balance are switched ON. The connections to the main grid are assumed to have sufficient capacity to exchange all the power needed by the clusters.

### 3.4. Blockchain implementation

To test the efficient operation of the blockchain network, a test bench has been built consisting of 10 single board computers (SBCs), namely Raspberry Pi model 4B [58]. Each cluster in the modified IEEE 123-node system (see Fig. 4) is emulated by two SBCs. The implementation of the BC is based on the Hyperledger Fabric framework [42] for the following reasons:

- The possibility of creating a permissioned BC network, in which members have to be previously authorized to have access the BC network and their transactions are always associated with their member. Therefore, each member of the BC network is linked to a physical person.
- The use of a Raft consensus algorithm [59], which allows the BC network to properly operate, even though several of the nodes in the BC network, stop working.
- The modular and open-source structure, which allows the BC architecture to be designed, to satisfy the project requirements, i.e., for different operating systems or processors, and different constraints such as the speed at which the blocks are added to the BC, the block

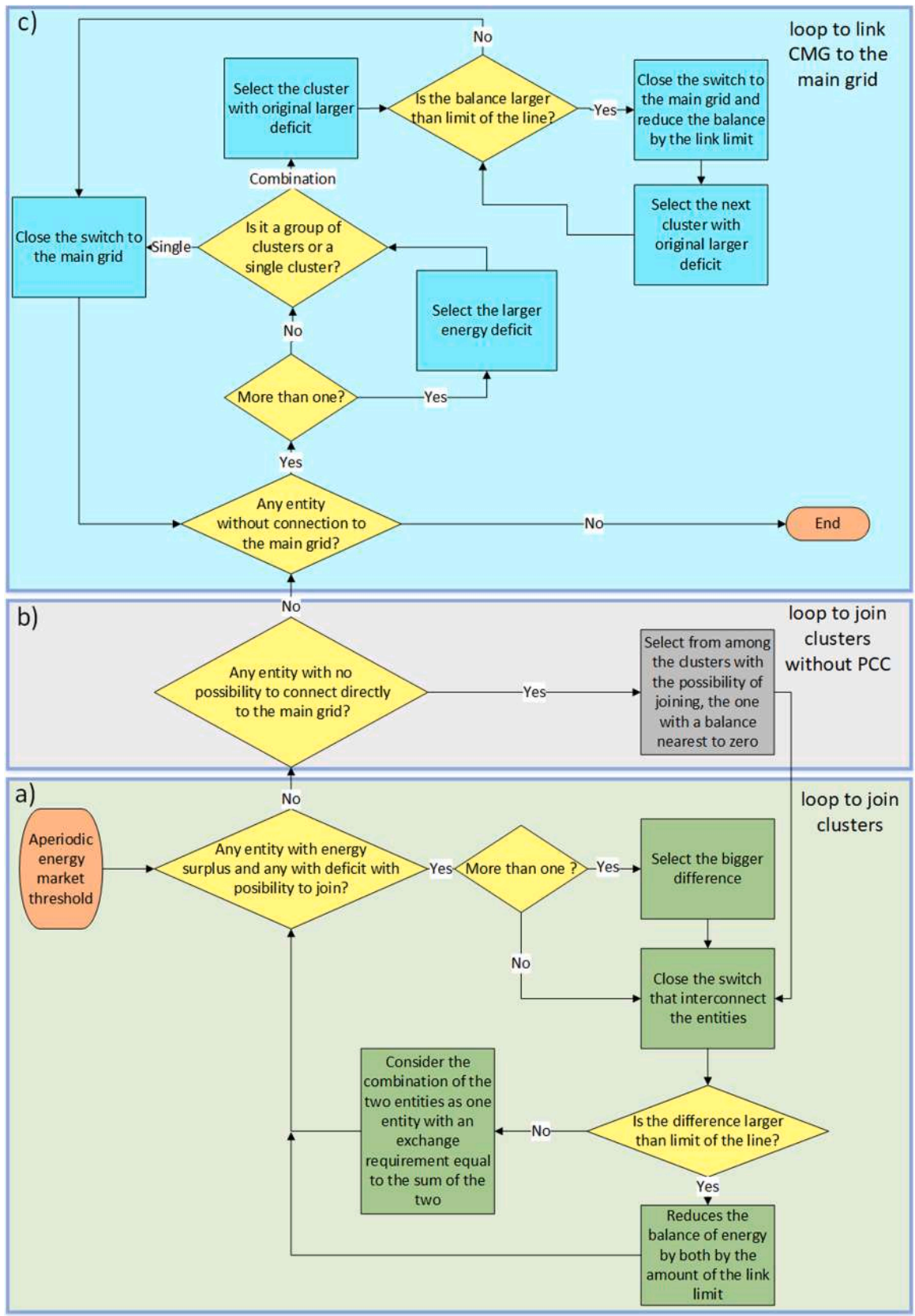

**Fig. 6.** Flowchart with the three loops of the proposed network reconfiguration algorithm: a) the first loop from bottom to top, joins the clusters to obtain clusters of clusters with an energy balance closer to zero; b) the second loop, joins clusters that do not belong to any other group of clusters yet and are not able to connect themselves to the main grid, trying to avoid that the balance of the new set significantly deviates from the zero balance; and c) the third loop joins each group of clusters to the main grid if necessary.

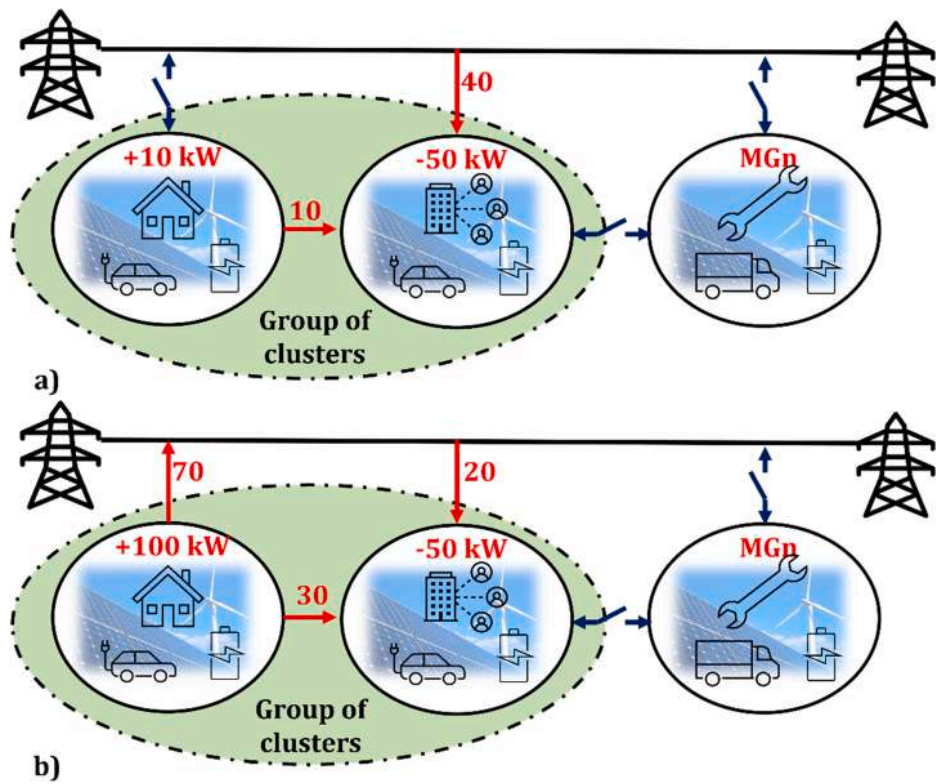

**Fig. 7.** Example of joining two clusters with a maximum link capacity of 30 kW.

size, the number of verifications before accepting each transaction, etc.

The modularity feature of the Hyperledger Fabric framework facilitates its implementation in the selected SBC, which paves the way for the use of low-cost, off-the-shelf devices. The modified version of the Hyperledger Fabric framework to run on ARM 64 processors and embedded libiec61850 library [60] has been published by the authors of this paper in a public repository [61], so that anyone can replicate the ARM-based BC network.

Fig. 8 shows, on the left, the code structure used to set up the blockchain network and the interaction with it and, on the right, an extract of the SC cluster, specifically the part that implements the reading of a user's Smart meter. This SC has been written in Go language, which is compatible with the C language, so that the libiec61850 library could be used without having to make any modifications to it.

### 4. Experimental and simulation results

This section presents the results obtained by emulating the proposed system. Firstly, the design and operation of the BC architecture based on Raspberry Pi 4 Model B boards is tested. SCs are implemented, and the times taken to return a result and the CPU usage are measured. Likewise, the operation of the LEM scheme and the network reconfiguration is verified. Secondly, the electrical behaviour is simulated in the standardized circuit presented in section 3.1 using the MATLAB-Simulink tool. The delays derived from those obtained from the experimental results are included to create as realistic a scenario as possible. The simulation has been carried out for two use cases, a weekday, and a weekend day. For each of the use cases, four scenarios have been evaluated that reflect the shift from the traditional network to the proposal of this work with management of the CMG topology. Thus, the improvement of system of the LEMs and the network reconfiguration management strategy can be quantified in terms of energy and economic parameters for both individual members and the whole of the CMG.

In this simulation scenario, it is assumed that the CMG members have neither storage capacity nor load regulation. This represents the most limited scenario from a control and economic point of view, since the CMG members cannot modify its load profiles in order to achieve better prices.

In addition, a comparison between the execution of the LEM on a periodic basis and the two aperiodic strategies described in section 2.2 has been carried out for the two use cases presented, weekday and weekend day.

### 4.1. Blockchain implementation

The test bench has been validated for 48 h without reporting any errors or causing any significant delay, which proves the feasibility of implementing a BC network in low-cost off-the-shelf devices based on the Hyperledger Fabric framework.

Fig. 9 shows a screenshot of the orderer logs displaying the start of the BC operation and how the leader, who will determine the order in which the transactions will be entered into the BC blocks, is elected. The same figure shows the writing of the first blocks.

Fig. 10 depicts another log screenshot, in this case of the peer, which at the beginning shows the nodes which form the BC network at that moment. It is also noted that this node is joined to the two DBs of which it must be a part, namely CMG DB and cluster DB. In order to perform these joins, a series of transactions have been created and stored in BC blocks.

Table 1. shows the parameters of interest regarding the tests performed in the SBC:

The CPU usage of each SBC ranges from 4.27 % to 10.02 % of the total computational capacity of the SBC, which means that the SBC computational power is capable to meet the computing requirements for proper operation.

The delay, defined as the execution delay of the different SCs, including the communications, varies between 0.7 and 24 s, with a mean of 1.8 s. To accurately simulate the case study, the worst-case delay

```go
package main

/*

#cgo LDFLAGS: -L /usr/lib -liec61850 -Wl,-rpath=/usr/lib
#include "iec61850_model.h"
#include "iec61850_client.h"

#include <stdlib.h>
#include <stdio.h>

#include "hal_thread.h"

float read() {
    char* hostname ="192.168.0.2";
    int tcpPort = 102;
    IedClientError error;
    IedConnection con = IedConnection_create();
    IedConnection_connect(con, &error, hostname, tcpPort);
    if (error == IED_ERROR_OK){
        MmsValue* energy_consumption =
            IedConnection_readObject(con, &error,
            "SMARTMTR/MMXU.Watt.mag.f", IEC61850_FC_MX);
        if (error == IED_ERROR_OK){
            float consumption =
            MmsValue_toFloat(energy_consumption);
            IedConnection_destroy(con);
            return consumption;
        }
    }
    IedConnection_destroy(con);
}

*/
import "C"

import (
    "encoding/json"
    "fmt"
    "strconv"

    "github.com/hyperledger/fabric-contract-api-go/contractapi"
)

// SmartContract provides functions for managing a user
type SmartContract struct {
    contractapi.Contract
}

// User describes basic details of what makes up a user
type User struct {
    Battery     float32 `json:"battery"`
    Generation  float32 `json:"generation"`
    Consumption float32 `json:"consumption"`
}
```

**Fig. 8.** Screenshot of the project code structure with details of the cluster SC showing how smart meter reading via MMS messages has been implemented following the IEC 61850 communications standard.

**Fig. 9.** Logs of the Docker orderer in operation showing how the leader node is chosen and the beginning blocks of the BC are written.

scenario of 24 s is considered. Thus, for each round of CMG energy market, 5 functions from 2 SCs must be executed (CEM, RUM, CMGEM, NR and RCM functions, explained in section 2.1). This results in a delay of 120 s for the worst-case scenario.

Fig. 11 shows a screenshot that displays the user interface containing the state of the world of the DB, i.e., the current value of each parameter, disregarding previous values. This interface is accessible via a web browser, through the port that has been determined in the configuration.

The interface is implemented in the Apache CouchDB [62], it has not been deployed by the authors.

Each BC block is composed of a header and one or more transactions. The number of transactions in each block is limited to 10 transactions or 1 s from the start of the block, whichever comes first. From the experimental results it has been obtained that the storage required for each BC block ranges from 4 to 32 kB, depending on the number of transactions the block contains i.e., the memory size required for each transaction ranges from 3.2 kB to 4 kB. Likewise, the storage capacity required by each of the nodes has been estimated considering as a transaction, each execution of a function of a SC. In the worst-case storage scenario, if each block contains only one transaction, each transaction requires a storage space of 4 kB. The memory taken up by the execution of the cluster energy markets depends on the number of members in the market, since one bid must be submitted to the market for each member. In addition, two functions must always be executed, regardless of the number of members: the cluster energy market (CEM) and the reading of users' meters (RUM). The memory size occupied by each market round can be estimated according to the following equation:

$$size_{round} = 2Å·size_{tx}(MAX) + NÅ·size_{tx}(MAX). \quad (5)$$

**Fig. 10.** Docker peer logs showing the peers which are currently running on the blockchain network, and the addition of the first blocks to the blockchain received from the orderer.

**Table 1**
Results obtained on the test bench.

|  | Min | Mean | Max |
|---|---|---|---|
| CPU usage (%) | 4.27 | 6.34 | 10.02 |
| Delay (s) | 0.7 | 1.8 | 24 |
| Storage (kB) | 3.2 | 3.4 | 4 |

where $size_{round}$ represents the memory size required by each cluster energy market round, $size_{tx}(MAX)$ is the storage required by each transaction, and N is the number of members participating in the cluster energy market. For each round of the cluster energy market, each member executes a function to submit a bid into the market then, another function is automatically executed to match the bids received and after that another function is automatically executed to read the members' smart meters and verify the actual consumption of the previous round.

The memory size required by the execution of the CMG energy market also depends on the number of participating clusters, since the execution of each function for the cluster energy market matching, results in the amount of energy intended to be exchanged in the market between clusters. This data must be stored in the DB of the CMG. In addition, three functions must always be executed, regardless of the number of clusters: the CMG energy market, the network reconfiguration and the reading of CMG's meters. Therefore, each market round within the CMG takes up a memory space which can be estimated according to the following formula:

$$size_{round} = 3 \cdot size_{tx}(MAX) + M \cdot size_{tx}(MAX) \quad (6).$$

where M is the number of clusters participating in the CMG market. For each CMG energy market round the following function executions take place: (i) one per cluster to submit the data of its cluster energy market round: (ii) one SC execution to match the bids of the clusters; (iii) another SC execution to obtain the optimal topology for the CMG; (iv) and a last transaction to read the PCC smart meters and the connection points between clusters.

### 4.2. Scenario comparison

To demonstrate the benefits of switch management at CMG, several scenarios are considered for the weekday use case. A comparative

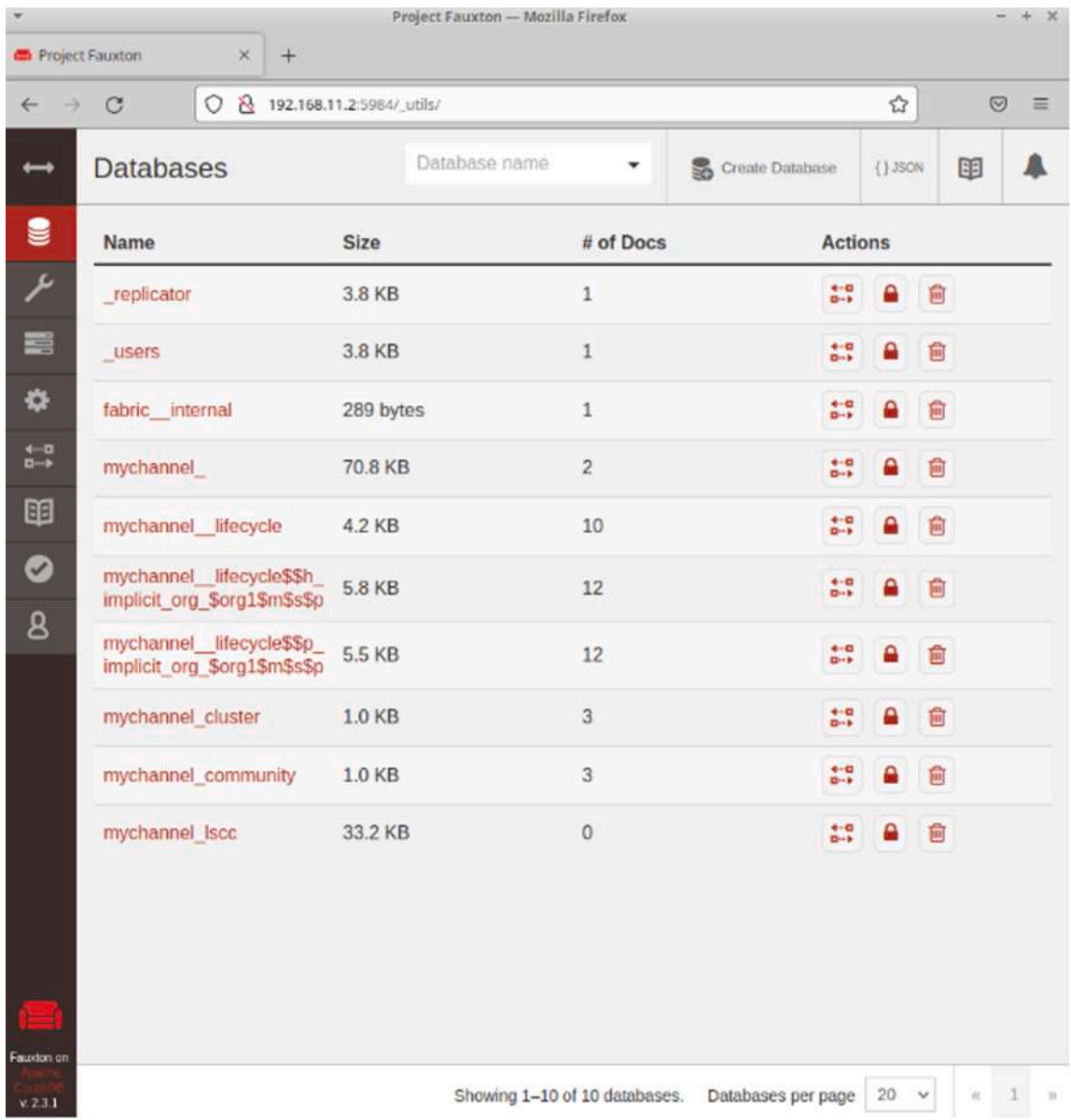

**Fig. 11.** Screenshot of the state of the world database that can be accessed through the web browser.

analysis is carried out to determine the scenario that leads to the minimum amount of energy injected to the main grid, resulting in lower operating costs since energy generated in the CMG is consumed within the CMG. The scenarios range from the traditional grid approach, in which each member acts as an independent entity, to the scenario in which a CMG with several clusters interconnected and control over network topology is considered, to take advantage of the differences in their generation and consumption profiles. The different scenarios are explained as follows:

**First scenario**: this scenario represents the current conventional power system in which each member can independently connect to the main grid. The DSO sets the price for the energy exchanges with the members.

**Second scenario**: in this scenario, each of the clusters proposed in section 3.1 is considered as a CMG which operates independently from the other clusters. In this scenario, which has become a reality nowadays, an aggregation of grid users in geographical proximity try to obtain better energy prices. There can be energy exchanges among the members within a particular CMG. Due to the restrictions imposed on cluster 2, which has no connection to the main grid, cluster 2 and cluster 5 are aggregated, so that the resulting cluster can maintain the frequency and voltage values within the safe margins. The different electrical profiles as a function of the role of each cluster are of no value. Therefore, a cluster can be injecting energy to the main grid whereas others can be drawing energy from the grid, at the price set by the DSO.

**Third scenario**: All the clusters are combined to form a unique CMG. This is a real scenario in which several CMGs, such as those shown in the second scenario, join together to try to take advantage of their different consumption and generation profiles and obtain better prices. With no control over the switches and with several PCCs, it is possible that a cluster is selling energy to the grid whereas another is purchasing, which does not create an ideal scenario for the members in terms of energy prices. Fig. 12 shows that the amount of energy injected by the whole CMG (in red) to the main grid is higher than the energy injections to the main grid, required to maintain the balance (in blue).

**Fourth scenario**: all the clusters are joined together to form a unique

CMG with control over the switches. The third scenario is modified by including switches in the PCCs and in the connections among clusters. Thus, clusters with complementary energy profiles are aggregated and the cluster through which energy is absorbed or injected to the main grid is controlled trying to avoid energy injections to the main grid. A market round among clusters is periodically made. The period is set to 6 min. This time is sufficient for the proper operation of the network and ensures compliance with the maximum time delays measured experimentally. Although a market round is periodically performed periodically, this does not imply that the network topology is periodically modified. The network reconfiguration only takes place when the market outcomes significantly change.

Once the different scenarios have been described, the circuit shown in Fig. 4 is simulated using the MATLAB-Simulink tool. A role is assigned to each node according to the cluster they belong to. The load profiles of each cluster are shown in Fig. 5. The SCs described in sections 3.1 and 3.2 have been implemented by using a MATLAB code script. To make the simulation more realistic, a delay has been introduced by a normal distribution with mean 120 s, the worst-case scenario of those obtained in the BC test of section 4.1, and a standard deviation of 60 s. This margin represents the time it takes for the system to receive updated data in order to start acting on the switches.

The MATLAB-Simulink simulation tool returns the currents and voltages in each of the cables that make up the IEEE 123-node system network. From these values, the energy exchanged with the main grid and the energy dissipated in the cables due to the joule effect are calculated (see equation (7)).

In scenario 1 it is assumed that the losses due to the energy transportation to the meter of each user are assumed by the DSO. On the other hand, in the other scenarios, the losses assumed by the DSO are up to the CMG connection point. Therefore, users must assume the energy losses from this connection point between the main grid and the CMG to their meters. To demonstrate that these energy losses to be assumed by the users are negligible compared to the amount of energy they no longer exchange with the main grid, these losses are calculated.

$$\sum_{i=0}^{i=n} Losses = R_i \cdot I_i^2 \cdot t \quad (7).$$

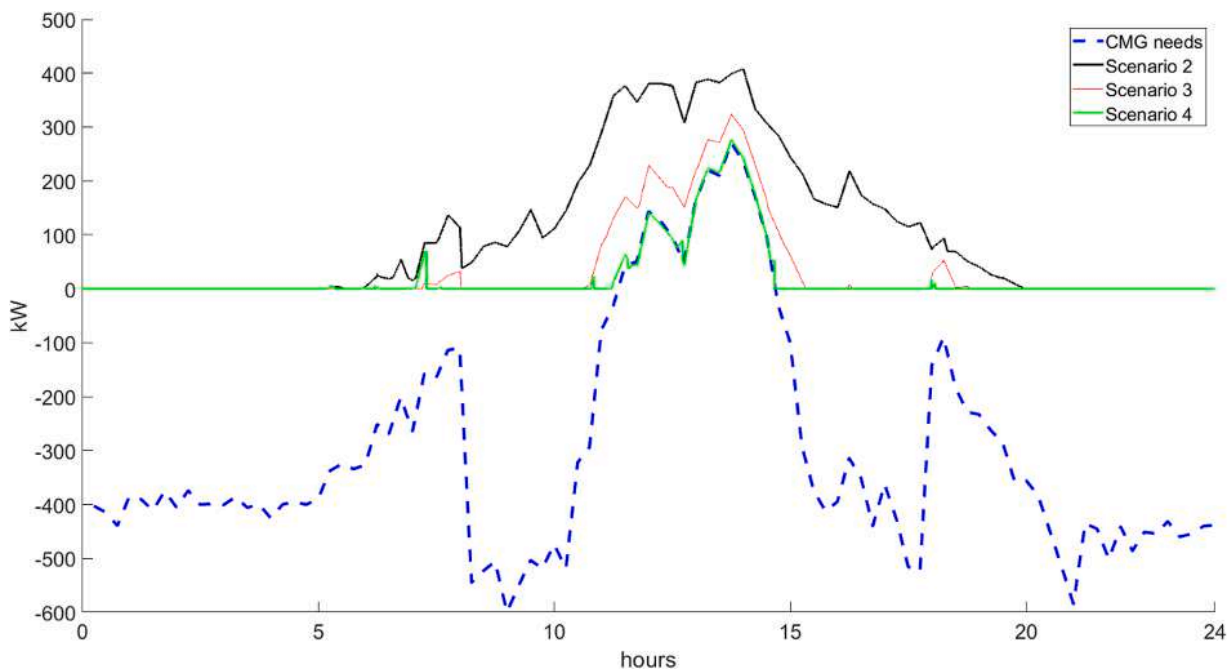

**Fig. 12.** The sum of the energy profiles of all CMG members throughout the day is shown in blue. The black line represents the energy injections to the main grid when each cluster is connected independently to the main grid. The red curve depicts the energy injections to the main grid when all clusters are merged into a CMG but with no control over the switches. The green line represents the case where there is control over the switches of the CMG. (For interpretation of the references to colour in this figure legend, the reader is referred to the web version of this article.)

Where $I_i$ is the current flowing through the cable, t is the time and $R_i$ is the resistance of the cable calculated by the following equation:

$R_i = \rho_i · S_i · l_i (8).$

Where $\rho$ is the cable resistivity, $S_i$ is the cable cross-section and $l_i$ is the cable length. All these parameters, for each of the cables, are defined in the IEEE 123-node system network.

The simulation results are depicted in Table 2 and graphically represented in Fig. 12. For a better understanding, a comprehensive comparative analysis of the transitions between case studies is carried out.

**Comparison of scenarios 1 and 2.**

The difference between the first and the second scenario is that users of the traditional grid are aggregated and become members of a CMG. Therefore, they can take advantage of their differences in energy consumption and generation profiles. The negative point is that they become responsible for the energy losses occurring in the distribution network. The savings are achieved by taking advantage of wide differential existing in most of the electricity markets worldwide between the price of buying and selling energy for members, as in the case of the Spanish electricity market, described in model [30]. Savings due to the price difference are much higher than the costs associated with energy losses. The results are conclusive since by grouping members to form the CMGs, the total energy exchanged with the main grid is reduced by 59 %. The losses caused in the energy transmission are totally negligible compared to the savings resulting from the shift from scenario 1 to scenario 2.

**Comparison of scenarios 2 and 3.**

There is a larger difference between scenario 2 and scenario 3 since a connection has been added among the different clusters. As a result, that they can take advantage of the different energy exchange profiles with the grid due to their different uses. In scenario 3, the energy exchange with the main grid is reduced by 25 % compared to scenario 2. In addition, by joining the clusters together, energy losses are also minimized, so that the total energy consumption is lower than that when the clusters individually operate. Shifting from scenario 2 to scenario 3, requires to physically link together clusters located in close proximity. Therefore, an investment must be made.

Fig. 12 shows the energy profile required by the CMG (dashed blue line), considering the positive sign as energy excess, and the real energy exchanges with the main grid (black line) that take place in scenario 2 and those occurring in scenario 3 (red line).

**Comparison of scenarios 3 and 4.**

The shift from scenario 3 and scenario 4 involves adding a controllable switch to each of the connections between clusters and the PCCs. These switches allow the clusters exchanging energy and the points where the CMG connect to the main grid to be selected. Implementation of this strategy results in a decrease in the amount of energy exchanged of 6.89 %. This is due to the fact that injections to the main grid have been cut from 846.51 kWh in scenario 3 to 515.46 kWh in scenario 4, for the minimum possible of 485.42 kWh. Fig. 12 shows how the green line, representing the injections to the main grid in scenario 4, perfectly fits with the CMG requirements (dashed blue line), in contrast to the red line which represents the injections to the main grid performed in scenario 3.

### 4.3. Weekend day use case

To confirm that the excellent results obtained do not depend on the use case under consideration, but improve the results in any scenario with energy surplus, the energy consumption profiles of the five clusters have been modified to simulate a weekend use case as follows:1) consumption in the industrial area has been reduced by 10 %; 2) the university campus is kept with a reduced consumption state throughout the day; 3) the commercial area increases its energy consumption during business hours by 100 %; and 4) both residential areas have increased their consumption by 50 %. The results obtained for the same scenarios as those described for the weekday use case are shown in Table 3.

The results shown in Table 3 for the second use case are similar to those obtained in the first use case, validating the strategy implemented. It can be seen that there is a decrease in unwanted energy injections to the main grid, resulting in a reduction in the total energy exchanged. As in the previous use case, the losses are also negligible when compared to the decrease in the amount of energy exchanged with the main grid.

Fig. 13 shows that, for the weekend day use case, the improvement yielded by the topology control in terms of reducing the amount of energy injected to the main grid becomes even more evident. If each of the clusters were operating independently, energy would be injected into the main grid between 5 and 20 h uninterruptedly (black line). By being able to reconfigure the network, energy is only injected to the grid for a few moments during the central hours of the day.

### 4.4. Aperiodic energy market

This subsection analyses the relevance of moving from a periodic market, as presented in the previous subsections, to an aperiodic market. The implementation of the aperiodic market has been achieved by using the techniques described in section 2.2 to detect when the power or energy variations are sufficiently relevant to trigger the aperiodic market: 1) SoD, which compares the power measured by the meters with that of the last result of the CMG market; 2) SoA, which evaluates the deviation in the amount of energy since the last result of the CMG market. Fig. 14 depicts the results obtained with each technique, SoD or SoA, with different thresholds, in the figure are presented: the number of changes in the topology of the CMG and the excess of energy fed into main grid compared to the periodic case presented in the previous subsections (Table 2 and Table 3).

From equations (1) and (3) it can be observed that for smaller thresholds the number of updates will generally be higher because the deviation that occurs between the current measurement and the last measurement sent to the Market is smaller, while the opposite is true for larger thresholds. For the weekday use case be noted that, as the SoD threshold increases, fewer updates are needed, with an increase of only 30 kWh regarding the minimum injections that are mandatory in the daily injections for 15 changes in the topology. As far as SoA is concerned, to reach similar figures in terms of number of topology changes, an excess of more than 80 kWh per day would be injected into the CMG.

The use of aperiodic techniques to establish the schedule of market rounds leads to significant advantages. Firstly, there is a saving in the amount of memory required for members' devices on account of the fewer market rounds required, and secondly, there is a reduction in the number of communications to be carried out among members, while maintaining a system performance similar to the periodic approach.

**Table 2**
Simulation results for the given scenarios.

| | Scenario 1 | Scenario 2 | Scenario 3 | Scenario 4 |
|---|---|---|---|---|
| Energy fed into grid (kWh) | 12154.3 | 2471.11 | 846.51 | 485.46 |
| Energy absorbed from grid (kWh) | 19618.85 | 10531.8 | 8853.02 | 8545.34 |
| Total internal CMG losses (kWh) | – | 39.29 | 22.87 | 36.34 |
| Total energy exchanged (kWh) | 31773.15 | 13002.91 | 9699.53 | 9030.8 |

**Table 3**
Simulation results for the given scenarios on a weekend day.

| | Scenario 1 | Scenario 2 | Scenario 3 | Scenario 4 |
|---|---|---|---|---|
| Energy fed into grid (kWh) | 14146.04 | 2226.67 | 1136.03 | 234.51 |
| Energy absorbed from grid (kWh) | 28799.67 | 18132.65 | 16966.24 | 16083.23 |
| Total internal CMG losses (kWh) | – | 166.48 | 74.33 | 158.21 |
| Total energy exchanged (kWh) | 42945.71 | 20359.32 | 18102.27 | 16,317 |

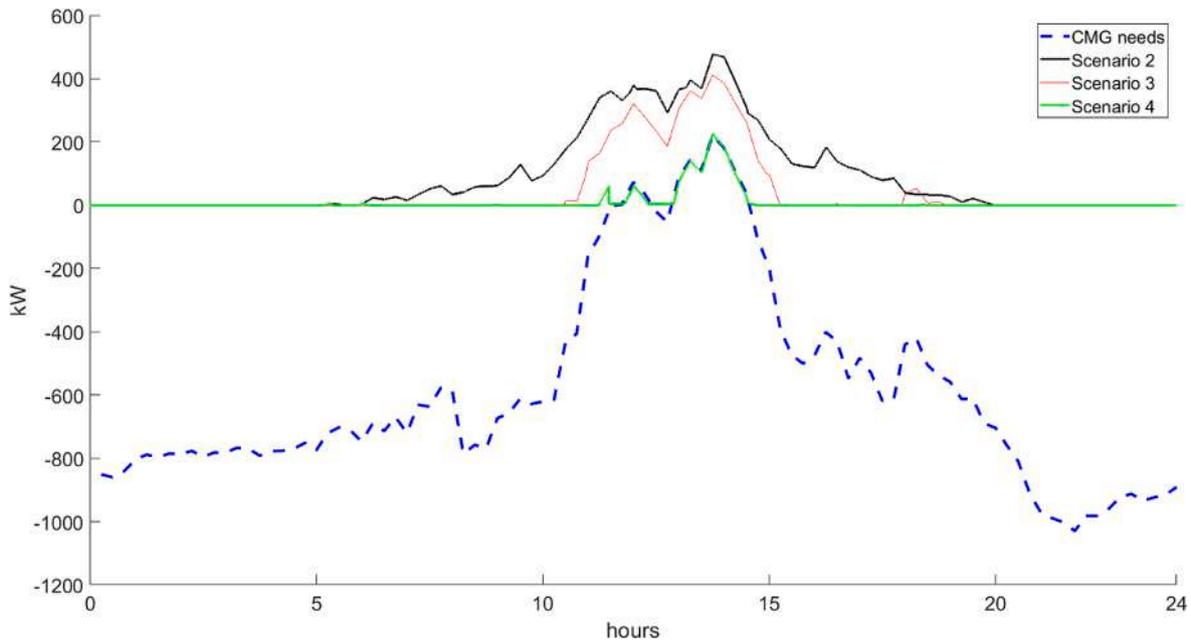

**Fig. 13.** The sum of the energy profiles of all CMG members throughout the weekend day is shown in blue. The black line represents the energy injected to the main grid when each cluster is connected independently to the main grid. The red curve depicts the energy injections to the main grid when all the clusters are merged into a CMG but with no control over the topology. The green line represents the case where there is control over the topology of the CMG. (For interpretation of the references to colour in this figure legend, the reader is referred to the web version of this article.)

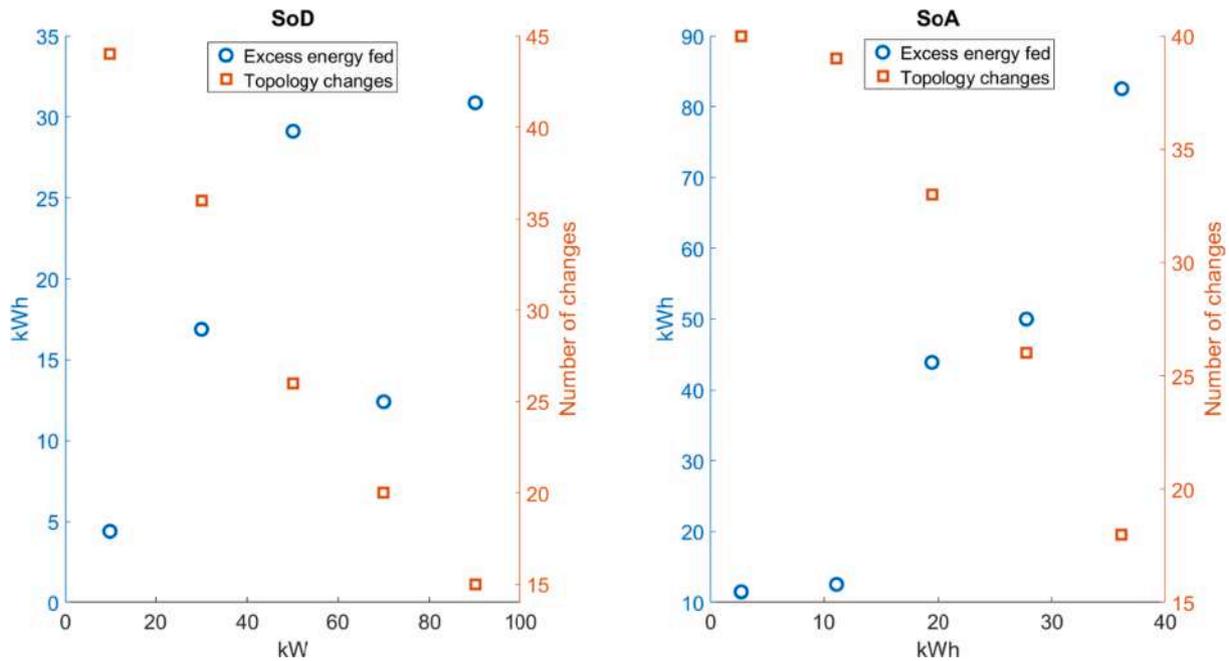

**Fig. 14.** Excess of energy fed into the main grid regarding the minimum injections that are mandatory and the number of changes in the in network topology of the CMG are presented for different threshold of aperiodic market for weekday use case.

Table 4. shows the reduction in the number of market rounds achieved by the aperiodic SoD technique in the case of the weekday, from 240 market rounds in the periodic technique every 6 min to 46 in the aperiodic technique without increasing the amount of energy exchanged with the grid.

This reduction in the number of market rounds has a positive impact on two aspects: 1) the reduced number of changes in the topology required increases the useful life of the switching elements 2) the amount of information to be stored by the DBs is also reduced.

Reducing the number of market rounds from 240 to 46 also results in a reduction of more than 80 % in the storage capacity requirements.

Similar results can be seen in Table 5 for the case of the weekend day, with a significant reduction in the number of market rounds by using aperiodic techniques.

**Table 4**

Comparison between the application of periodic control and different thresholds for the two aperiodic control techniques used in the weekday.

|  | Periodic | Aperiodic SoD | | | Aperiodic SoA | | |
|---|---|---|---|---|---|---|---|
|  | Ts = 6 min | $\Delta_{SoD}$ = 10 kW | $\Delta_{SoD}$ = 30 kW | $\Delta_{SoD}$ = 50 kW | $\Delta_{SoA}$ = 11 kWh | $\Delta_{SoA}$ = 20 kWh | $\Delta_{SoA}$ = 28 kWh |
| Energy fed into grid (kWh) | 485.42 | 488.08 | 502.29 | 514.53 | 497.96 | 529.32 | 535.43 |
| Energy absorbed from grid (kWh) | 8545.34 | 8546.67 | 8547.34 | 8566.45 | 8545.59 | 8581.56 | 8580.63 |
| Total internal CMG losses (kWh) | 36.34 | 35.71 | 35.02 | 35.22 | 34.72 | 35.75 | 34.90 |
| Total energy exchanged (kWh) | 9030.8 | 9034.75 | 9049.63 | 9080.97 | 9043.55 | 9110.88 | 9116.06 |
| # CMG markets | 240 | 221 | 83 | 46 | 156 | 89 | 62 |
| Changes of topology | 35 | 44 | 36 | 26 | 39 | 33 | 26 |

**Table 5**

Comparison between the application of periodic control and different thresholds for the two aperiodic control techniques used in the weekend day.

|  | Periodic | Aperiodic SoD | | | Aperiodic SoA | | |
|---|---|---|---|---|---|---|---|
|  | Ts = 6 min | $\Delta_{SoD}$ = 10 kW | $\Delta_{SoD}$ = 30 kW | $\Delta_{SoD}$ = 50 kW | $\Delta_{SoA}$ = 11 kWh | $\Delta_{SoA}$ = 20 kWh | $\Delta_{SoA}$ = 28 kWh |
| Energy fed into grid (kWh) | 234.51 | 235.78 | 233.17 | 243.59 | 236.93 | 234.51 | 233.33 |
| Energy absorbed from grid (kWh) | 16083.23 | 16083.29 | 16076.57 | 16092.20 | 16083.55 | 16144.43 | 16082.25 |
| Total internal CMG losses (kWh) | 158.21 | 158.71 | 158.76 | 158.85 | 158.83 | 162.69 | 158.02 |
| Total energy exchanged (kWh) | 16317.75 | 16319.08 | 16309.74 | 16335.80 | 16320.48 | 16378.94 | 16315.58 |
| # CMG markets | 240 | 232 | 93 | 52 | 178 | 164 | 143 |
| Changes of topology | 26 | 26 | 16 | 14 | 22 | 22 | 22 |

### 4.5. Analysis of results

From the results obtained by evaluating the two selected use cases (weekday and weekend day) it can be concluded that the reconfiguration of the distribution network minimises the exchange energy between the CMG and the main grid. This results in lower energy costs for CMG members which are able to use the energy that is generated within the CMG and thereby reducing the amount of energy drawn from the main grid.

The discussion continues analysing the aperiodic solution. Then a study of technical feasibility is carried out. Finally, the consequences of the delays are explored.

#### 4.5.1. Aperiodic solution

The SoD aperiodic strategy with a 30 kW threshold has been adopted as the most optimal solution for two reasons. First, it maintains the energy exchange with the main grid very similar to that obtained by the periodic solution and, second, it significantly reduces the storage capacity required by decreasing the number of market rounds. It is important to note, however, that any of the solutions proposed in the previous section could be implemented. The choice between SoD and SoA is based on the control requirements the designer sets for the system. Fig. 15 shows the power transferred across the links of each cluster and its power balance (dashed red line), the positive vertical axis representing power injections and the negative one the power absorptions. From the figure, it can be observed that, in the central hours of the day, there is an energy surplus in clusters 1, 3 and 4 due to the fact that the photovoltaic installations are oversized. In the case of cluster 4, instead of injecting the energy surplus to the main grid, it is transferred to cluster 5, which has energy deficit during the central hours of the day, thereby achieving better energy prices for both parties involved in the energy exchange. Given that cluster 2 has no connection to the main grid, it can obtain energy from cluster 1 and cluster 5, depending on which of them best complements its demand curve. Cluster 1 shows the most complex case, because it produces a surplus of energy at some times of the day. However, cluster 1 must buy energy from the main grid or from cluster 3 since cluster 1 has to sell energy to cluster 2 so that the latter can maintain its power balance. In the central hours of the day, between cluster 4 and 5, there is a surplus of energy that cannot be used by the rest of the clusters, which have no energy demand at that moment. In this case, cluster 5 is used to connect with the main grid, preventing cluster 4, which is where this surplus is really being generated, from injecting more energy into the grid than it really has left over.

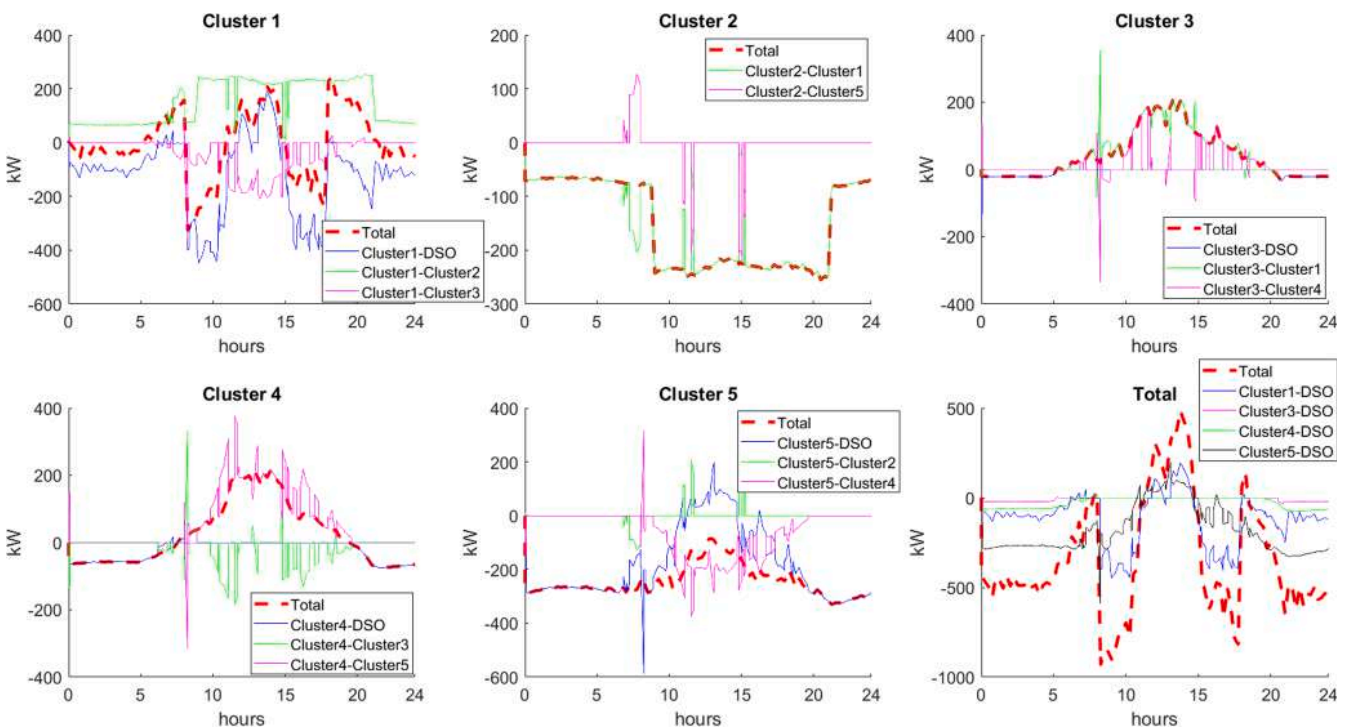

**Fig. 15.** Profiles of the power transmitted through the CMG links between clusters and with the main grid.

### 4.5.2. Study of the technical feasibility

To verify that the quality-of-service requirements are satisfied, voltages and frequencies have been measured at the PCCs. The results can be seen in Fig. 16. Regarding the frequency, the variations with respect to the main grid (ideal 60 Hz) are negligible, ranging from 59.9998 Hz to 60.002 Hz. As for the voltage, the deviation with respect to the main grid voltage is always below 7 %. These deviations take place when there is an excess of energy in the cluster and the connection to the main grid is closed so that this energy is absorbed by another cluster.

These results are in line with expectations. One of the convergence constraints of the network reconfiguration algorithm is that all CMG members must always have an available path to the main grid.

### 4.5.3. Delay analysis

The consequences of a longer delay in the acquisition of member data for the execution of the SC responsible for network reconfiguration, have been experimentally simulated. The excess of energy as a function of the communication delay can be seen in Table 6.

It can be seen that, as the delay is increased, a greater amount energy that could be used within the CMG will be injected into the main grid. In the unlikely event that the delay is, for instance, increased by 30 min, the excess energy injected into the main grid would be around 200 kWh over the simulated day, which is less than the amount of energy that would be injected if the switches were not controlled.

**Table 6**
Excess of energy fed into main grid for different communication delays.

| Delay (minutes) | Excess of energy fed into main grid (kWh) |
| --- | --- |
| 2 | 2.66 |
| 6 | 10.35 |
| 12 | 28.69 |
| 18 | 34.59 |
| 24 | 98.05 |
| 30 | 206.49 |

## 5. Conclusion

This paper proposes a new two-layer energy management strategy for CMGs which are organized in clusters of members. The first layer, referred as the market layer, deals with the creation of the LEMs to manage the energy exchange among the community members. The second layer, referred as reconfiguration layer, changes the network topology to reduce the amount of energy injected to the main grid. The primary purpose of this two-layer-based EMS is to maximise the economic benefits for the members of the entire local community by reducing the energy injections to the main grid, which in turn increases the energy independence and reduces energy losses since the need of energy from remote generation plants is reduced. This greatly facilitates the penetration of DERs. The implementation of the reconfiguration algorithm is based on SCs which allow libraries of functions for managing IEC 61850 compatible equipment and devices to be used. Therefore, the network can be reconfigured directly by running the SCs.

The market layer is divided into two sub-layers: the lower market layer, also called the cluster energy market and the upper layer, referred as CMG energy market. The former is used to create a LEM of each cluster and enables the first energy exchange between members within the cluster, whereas the latter addresses the energy exchanges between clusters once the preliminary stage of the cluster energy market has taken place. This division into sub-layers leads to a reduction in the computational resources required by the community members as only the details of the members comprising the corresponding cluster are processed disregarding the data from other clusters. As a result, the designed strategy can be implemented in low-cost off-the-shelf devices that play the role of BC nodes thereby facilitating the creation of CMGs with a small initial investment. This initial investment is further reduced by the applying the concept of aperiodic market triggering techniques which minimize the development costs associated with the storage capacity and communication channels, due to the reduction in number of interactions among the different nodes. The latter offers two clear advantages. Firstly, less information has to be stored in memory and secondly, the communication channels are free of redundant information.

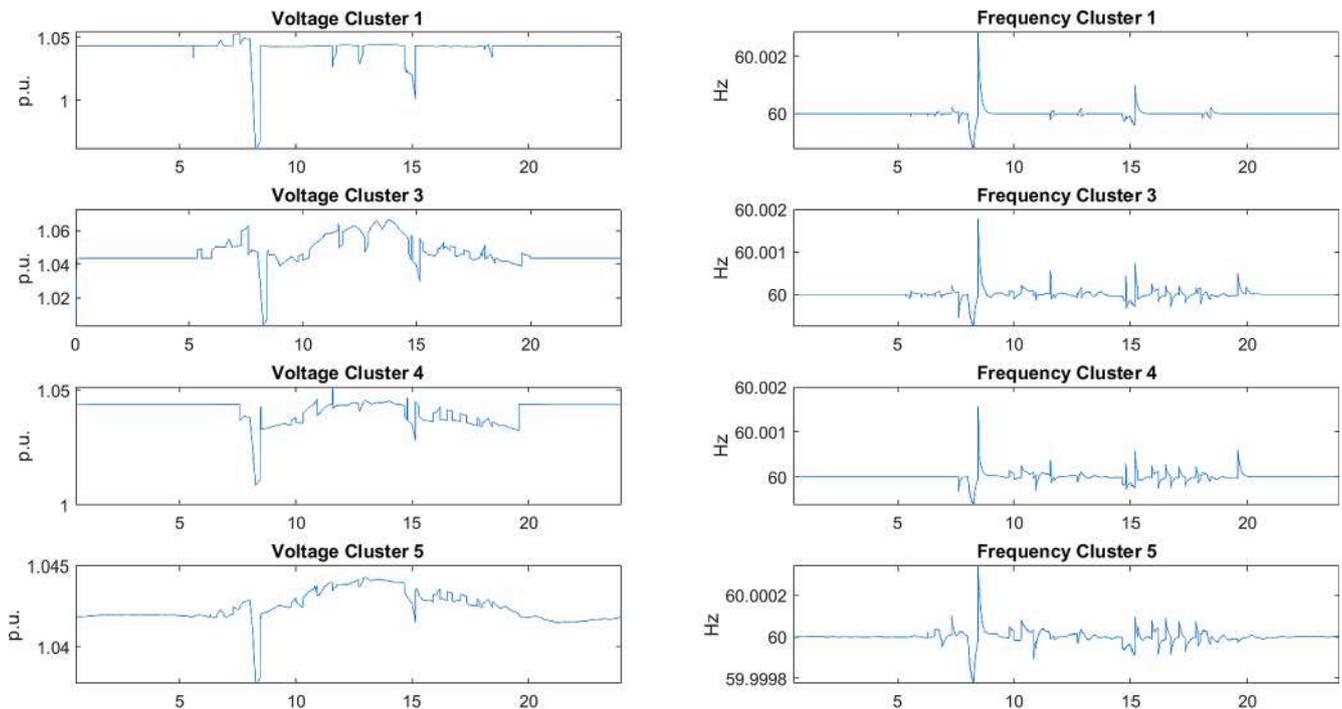

**Fig. 16.** Voltage and frequency at the connection points to the main grid of each of the clusters.

Two use cases are tested based on the IEEE 123-node system network model: one for a weekday and another for a weekend day. For each use case, several scenarios are considered, and a comparative analysis is performed to determine the scenario that leads to the minimum amount of energy injected to the main grid. Especially relevant is the shift from scenario 3 to scenario 4. The inclusion of controllable switches in a CMG, reduces energy costs by minimising energy exchange with main grid, e. g., for the weekday and the weekend day use cases the energy exchange is reduced by 6.89 % and 9.86 %, respectively. In addition, the use of aperiodic market techniques allows the amount of storage required to be significantly reduced, without a noticeable impact on the system performance. Using the SoD technique with a 30 kW threshold, the number of market rounds is reduced from 240 in the periodic case to 83 in the weekday use case and 93 in the weekend day use case. For the weekday use case, the increase observed in the energy exchanged with the main grid is only 18.83 kWh whereas for the weekend day use case the amount of energy exchanged is decreased by 8.01 kWh.

Although specific LEM strategies and a particular reconfiguration algorithm have been implemented to demonstrate the effectiveness of the energy management approach, the modularity concept employed in the SCs allows each CMG and even each cluster within the CMG to use the LEM mechanisms and reconfiguration algorithms that best suit their characteristics. Likewise, other targets can be reached, such as those related to ancillary services, voltage, and frequency support, etc. [63]. To that end, it would only require to implement new control strategies in the SCs and add a new member to represent the DSO in the LEM.

As far as the limitations are concerned, the shift from scenario 1 to scenario 4 must address both technical and social aspects. CMGs are technically complex, and the authors believe that the strategy developed in this paper facilitates the implementation and operation of CMGs which reduces social rejection. Gaining general acceptance by potential customers is crucial for any transformation of the energy system in the shape of CMGs to act as aggregators. In this regard, the DSOs and governments alike stand out as critical actors. Tight regulations and legal barriers hinder the widespread penetration of CMGs.

The next research step will be the installation of the BC node in a test MG to corroborate the correct functioning of the concept demonstrated in simulation. Future work should take into account the growing trend in installing small energy storage systems and the increasing penetration of electric vehicles, which may have a triple role in their interaction with the grid, consuming, injecting and storing energy.

## CRediT authorship contribution statement

**Miguel Gayo-Abeleira:** Conceptualization, Methodology, Validation, Formal analysis, Investigation, Data curation, Writing – original draft, Visualization. **Carlos Santos:** Conceptualization, Methodology, Validation, Writing – review & editing, Supervision. **Francisco Javier Rodríguez Sánchez:** Conceptualization, Methodology, Validation, Writing – review & editing, Supervision, Project administration, Funding acquisition. **Pedro Martín:** Conceptualization, Methodology, Validation, Writing – review & editing, Visualization, Supervision. **José Antonio Jiménez:** Supervision, Resources. **Enrique Santiso:** Supervision, Methodology.

## Declaration of Competing Interest

The authors declare that they have no known competing financial interests or personal relationships that could have appeared to influence the work reported in this paper.

## Data availability

Data will be made available on request.

## Acknowledgements

This work was supported in part by the PROMINT-CM Project funded by the Comunidad de Madrid and the European Social Fund under Grant P2018/EMT4366, and in part by the COPILOT-CM Project by the Comunidad de Madrid under Grant Y2020/EMT-6368.

## References

[1] F. Martins, C. Felgueiras, M. Smitkova y N. Caetano, Analysis of Fossil Fuel Energy Consumption and Environmental Impacts in European Countries, *Energies*, vol. 12, p. 964, March 2019.

[2] E. C. D. G. for Energy., Clean energy for all Europeans., Publications Office, 2019.

[3] Centre ECJR. Energy communities: an overview of energy and social innovation., Publications. Office 2020.

[4] J. Lowitzsch, C. E. Hoicka y F. J. van Tulder, Renewable energy communities under the 2019 European Clean Energy Package – Governance model for the energy clusters of the future?, *Renewable and Sustainable Energy Reviews*, vol. 122, p. 109489, 2020.

[5] D. Frieden, A. Tuerk, J. Roberts, S. d'Herbemont y A. Gubina, Collective self-consumption and energy communities: Overview of emerging regulatory approaches in Europe, Compile, 2019.

[6] R. H. Lasseter y P. Paigi, Microgrid: a conceptual solution, de *2004 IEEE 35th Annual Power Electronics Specialists Conference (IEEE Cat. No.04CH37551)*.

[7] M. Warneryd, M. Håkansson y K. Karltorp, Unpacking the complexity of community microgrids: A review of institutions' roles for development of microgrids, vol. 121, p. 109690, 4 2020.

[8] Coalition C. Community microgrids 2015.

[9] E. M. Gui, M. Diesendorf y I. MacGill, Distributed energy infrastructure paradigm: Community microgrids in a new institutional economics context, *Renewable and Sustainable Energy Reviews*, vol. 72, p. 1355–1365, 5 2017.

[10] Karavas C-S, Kyriakarakos G, Arvanitis KG, Papadakis G. A multi-agent decentralized energy management system based on distributed intelligence for the design and control of autonomous polygeneration microgrids. Energy Convers Manage 2015;103:166–79.

[11] V. Boglou, C.-S. Karavas, K. Arvanitis y A. Karlis, A Fuzzy Energy Management Strategy for the Coordination of Electric Vehicle Charging in Low Voltage Distribution Grids, *Energies*, vol. 13, p. 3709, July 2020.

[12] Cornélusse B, Savelli I, Paoletti S, Giannitrapani A, Vicino A. A community microgrid architecture with an internal local market. Appl Energy 2019;242: 547–60.

[13] Hvelplund F. Renewable energy and the need for local energy markets. Energy 2006;31:2293–302.

[14] O. Abrishambaf, F. Lezama, P. Faria y Z. Vale, Towards transactive energy systems: An analysis on current trends, *Energy Strategy Reviews*, vol. 26, p. 100418, November 2019.

[15] Y. Wu, Y. Wu, J. M. Guerrero y J. C. Vasquez, Decentralized transactive energy community in edge grid with positive buildings and interactive electric vehicles, *International Journal of Electrical Power {\&}amp$\mathsemicolon$ Energy Systems*, vol. 135, p. 107510, February 2022.

[16] S. Thakar, A. S. Vijay y S. Doolla, System reconfiguration in microgrids, *Sustainable Energy, Grids and Networks*, vol. 17, p. 100191, March 2019.

[17] Hemmatpour MH, Mohammadian M, Gharaveisi AA. Optimum islanded microgrid reconfiguration based on maximization of system loadability and minimization of power losses. Int J Electr Power Energy Syst 2016;78:343–55.

[18] M. I. Pathan, M. Al-Muhaini y S. Z. Djokic, Optimal reconfiguration and supply restoration of distribution networks with hybrid microgrids, *Electric Power Systems Research*, vol. 187, p. 106458, October 2020.

[19] Jabbari-Sabet R, Moghaddas-Tafreshi S-M, Mirhoseini S-S. Microgrid operation and management using probabilistic reconfiguration and unit commitment. Int J Electr Power Energy Syst 2016;75:328–36.

[20] Mengelkamp E, Notheisen B, Beer C, Dauer D, Weinhardt C. E. Mengelkamp, B. Notheisen, C. Beer, D. Dauer y C. Weinhardt, A blockchain-based smart grid. Computer Science - Research and Development 2018;33(1-2):207–14.

[21] Y. K. Renani, M. Ehsan y M. Shahidehpour, Optimal Transactive Market Operations With Distribution System Operators, *IEEE Transactions on Smart Grid*, vol. 9, p. 6692–6701, 11 2018.

[22] Shahidehpour M, Yan M, Shikhar P, Bahramirad S, Paaso A. Blockchain for Peer-to-Peer Transactive Energy Trading in Networked Microgrids: Providing an Effective and Decentralized Strategy. IEEE Electrific Mag 2020;8(4):80–90.

[23] M. Yan, M. Shahidehpour, A. Paaso, L. Zhang, A. Alabdulwahab y A. Abusorrah, Distribution Network-Constrained Optimization of Peer-to-Peer Transactive Energy Trading among Multi-Microgrids, *IEEE Transactions on Smart Grid*, vol. 12, p. 1033–1047, 3 2021.

[24] T. Capper, A. Gorbatcheva, M. A. Mustafa, M. Bahloul, J. M. Schwidtal, R. Chitchyan, M. Andoni, V. Robu, M. Montakhabi, I. J. Scott, C. Francis, T. Mbavarira, J. M. Espana y L. Kiesling, Peer-to-peer, community self-consumption, and transactive energy: A systematic literature review of local energy market models, *Renewable and Sustainable Energy Reviews*, vol. 162, p. 112403, July 2022.

[25] Mengelkamp E, Gärttner J, Rock K, Kessler S, Orsini L, Weinhardt C. Designing microgrid energy markets. Appl Energy 2018;210:870–80.

[26] J. Yang, J. Dai, H. B. Gooi, H. D. Nguyen y P. Wang, Hierarchical Blockchain Design for Distributed Control and Energy Trading Within Microgrids, *IEEE Transactions on Smart Grid*, vol. 13, p. 3133–3144, July 2022.

[27] Koirala BP, Koliou E, Friege J, Hakvoort RA, Herder PM. Energetic communities for community energy: A review of key issues and trends shaping integrated community energy systems. Renew Sustain Energy Rev 2016;56:722–44.

[28] A. Camilo Avilés, H. Sebastian Oliva y D. Watts, Single-dwelling and community renewable microgrids: Optimal sizing and energy management for new business models, *Applied Energy*, vol. 254, p. 113665, November 2019.

[29] S. Wang, A. F. Taha, J. Wang, K. Kvaternik y A. Hahn, Energy Crowdsourcing and Peer-to-Peer Energy Trading in Blockchain-Enabled Smart Grids, *IEEE Transactions on Systems, Man, and Cybernetics: Systems*, vol. 49, p. 1612–1623, August 2019.

[30] M. Gayo, C. Santos, F. J. R. Sanchez, P. Martin, J. A. Jimenez y M. Tradacete, Addressing Challenges in Prosumer-Based Microgrids With Blockchain and an IEC 61850-Based Communication Scheme, *IEEE Access*, vol. 8, p. 201806–201822, 2020.

[31] A. Arif, Z. Wang, J. Wang y C. Chen, Power Distribution System Outage Management With Co-Optimization of Repairs, Reconfiguration, and DG Dispatch, *IEEE Transactions on Smart Grid*, vol. 9, p. 4109–4118, September 2018.

[32] Yoldaş Y, Önen A, Muyeen SM, Vasilakos AV, Alan I. Y. Yoldaş, A. Önen, S. M. Muyeen, A. V. Vasilakos y İ. Alan, Enhancing smart grid with microgrids. Renew Sustain Energy Rev 2017;72:205–14.

[33] Huang W-T, Chen H-T, Chen H-T, Yang J-S, Lian K-L, Chang Y-R, et al. A Two-stage Optimal Network Reconfiguration Approach for Minimizing Energy Loss of Distribution Networks Using Particle Swarm Optimization Algorithm. Energies 2015;8:13894–910.

[34] I. Sanz, M. Moranchel, J. Moriano, F. J. Rodriguez y S. Fernandez, Reconfiguration Algorithm to Reduce Power Losses in Offshore HVDC Transmission Lines, *IEEE Transactions on Power Electronics*, vol. 33, p. 3034–3043, April 2018.

[35] An Integrated Approach. S. Tan, J.-X. Xu y S. K. Panda, Optimization of Distribution Network Incorporating Distributed Generators. IEEE Trans Power Syst 2013;28:2421–32.

[36] Zhan X, Xiang T, Chen H, Zhou Bo, Yang Z. Vulnerability assessment and reconfiguration of microgrid through search vector artificial physics optimization algorithm. International Journal of Electrical Power {&}amp$mathsemicolon$ Energy Systems 2014;62:679–88.

[37] M. M. A. Abdelaziz, H. E. Farag y E. F. El-Saadany, Optimum Reconfiguration of Droop-Controlled Islanded Microgrids, *IEEE Transactions on Power Systems*, vol. 31, p. 2144–2153, May 2016.

[38] Shariatzadeh F, Vellaithurai CB, Biswas SS, Zamora R, Srivastava AK. Real-Time Implementation of Intelligent Reconfiguration Algorithm for Microgrid. IEEE Trans Sustain Energy 2014;5(2):598–607.

[39] M. Khederzadeh, Enhancement of microgrid resiliency by mitigating cascading failures through reconfiguration, de *CIRED Workshop 2016*, 2016.

[40] Habiboost M, Bathaee SMT. A self-supporting approach to EV agent participation in smart grid. Int J Electr Power Energy Syst 2018;99:394–403.

[41] I. E. Commission, IEC 61850 Communication Networks and Systems in Substations, Geneva, 2002.

[42] S. Marzal, R. Salas, R. González-Medina, G. Garcerá y E. Figueres, Current challenges and future trends in the field of communication architectures for microgrids, *Renewable and Sustainable Energy Reviews*, vol. 82, p. 3610–3622, February 2018.

[43] T. S. Ustun, C. Ozansoy y A. Zayegh, Modeling of a Centralized Microgrid Protection System and Distributed Energy Resources According to IEC 61850-7-420, *IEEE Transactions on Power Systems*, vol. 27, p. 1560–1567, August 2012.

[44] R. E. Mackiewicz, Overview of IEC 61850 and benefits, de *2006 IEEE Power Engineering Society General Meeting*, 2006.

[45] Raspberry Pi, [Online]. Available: https://www.raspberrypi.com/products/raspberry-pi-4-model-b/.

[46] A. Gholami, F. Aminifar y M. Shahidehpour, Front Lines Against the Darkness: Enhancing the Resilience of the Electricity Grid Through Microgrid Facilities, *IEEE Electrification Magazine*, vol. 4, p. 18–24, March 2016.

[47] V. Mohan, J. G. Singh, W. Ongsakul y M. P. R. Suresh, Performance enhancement of online energy scheduling in a radial utility distribution microgrid, *International Journal of Electrical Power & Energy Systems*, vol. 79, p. 98–107, 7 2016.

[48] G. Liu, M. R. Starke, B. Ollis y Y. Xue, Networked microgrids scoping study, ORNL, TN.[Online]. Available: https://info. ornl. gov/sites/publications/files/Pub68339. pdf, pp. 0093–9994, 2016.

[49] J. Silvente y L. G. Papageorgiou, An MILP formulation for the optimal management of microgrids with task interruptions, *Applied Energy*, vol. 206, p. 1131–1146, 11 2017.

[50] Backhaus SN, Dobriansky L, Glover S, Liu C-C, Looney P, Mashayekh S, et al. Networked Microgrids Scoping Study. Office of Scientific and Technical Information (OSTI); 2016.

[51] Y. Liu, Y. Fang y J. Li, Interconnecting Microgrids via the Energy Router with Smart Energy Management, *Energies*, vol. 10, p. 1297, 8 2017.

[52] R. Tonkoski, D. Turcotte y T. H. M. El-Fouly, Impact of High PV Penetration on Voltage Profiles in Residential Neighborhoods, *IEEE Transactions on Sustainable Energy*, vol. 3, p. 518–527, 7 2012.

[53] L. Feng, H. Zhang, W.-T. Tsai y S. Sun, System architecture for high-performance permissioned blockchains, vol. 13, p. 1151–1165, 7 2019.

[54] Y. Jiang, K. Zhou, X. Lu y S. Yang, Electricity trading pricing among prosumers with game theory-based model in energy blockchain environment, *Applied Energy*, vol. 271, p. 115239, August 2020.

[55] Air Quality Monitoring in Smart Cities. C. Santos, J. A. Jimenez y F. Espinosa, Effect of Event-Based Sensing on IoT Node Power Efficiency. Case Study. IEEE Access 2019;7:132577–86.

[56] PES, IEEE PES Test Feeder, [En línea]. Available: https://cmte.ieee.org/pes-testfeeders/resources/. (Último acceso: 2021].

[57] Wei C. Generation and load data for IEEE 123-node case and Nanji case. IEEE DataPort 2019.

[58] R. Pi, Raspberry Pi 4 Model B, [En línea]. Available: https://www.raspberrypi.com/products/raspberry-pi-4-model-b/.

[59] D. Ongaro y J. Ousterhout, The raft consensus algorithm, *Lecture Notes CS*, vol. 190, 2015.

[60] MZ Automation, [En línea]. Available: https://libiec61850.com/.

[61] M. Gayo, Docker images Hyperledger Fabric ARM, Marzo 2020. [En línea]. Available: https://hub.docker.com/u/miguelgayo.

[62] T. A. S. Foundation, *CouchDB*.

[63] T. U. Solanke, V. K. Ramachandaramurthy, J. Y. Yong, J. Pasupuleti, P. Kasinathan y A. Rajagopalan, A review of strategic charging–discharging control of grid-connected electric vehicles, *Journal of Energy Storage*, vol. 28, p. 101193, April 2020.